\documentclass[reprint,amsmath,amssymb,aps]{revtex4-1}
\usepackage[colorlinks,citecolor=blue,linkcolor=blue,anchorcolor=blue,filecolor=blue,
urlcolor=blue]{hyperref}
\usepackage{graphicx}% Include figure files
\usepackage{ulem}
\usepackage{dcolumn}% Align table columns on decimal point
\usepackage{bm}% bold math
\usepackage{todonotes}

\begin{document}

\title{Dark matter effects on hybrid star properties}

\author{C. H. Lenzi$^1$, M. Dutra$^{1,2}$, O. Louren\c{c}o$^{1,2}$, L. L. Lopes$^3$, D. P. Menezes$^4$}
\affiliation{
\mbox{$^1$Departamento de F\'isica, Instituto Tecnol\'ogico de Aeron\'autica, DCTA, 12228-900, 
S\~ao Jos\'e dos Campos, SP, Brazil} \\ 
\mbox{$^2$Universit\'e de Lyon, Universit\'e Claude Bernard Lyon 1, CNRS/IN2P3, IP2I Lyon, UMR 5822, F-69622, Villeurbanne, France} \\
\mbox{$^3$ Centro Federal de Educa\c{c}\~ao Tecnol\'ogica de Minas Gerais, Campus VIII CEP 37.022-560, Varginha, MG, Brazil} \\
\mbox{$^4$ Depto de Física, CFM, Universidade Federal de Santa Catarina, Florianópolis, SC, CP:476, CEP 88.040-900, Brazil }
}

\begin{abstract}
In the present work we investigate the effects of dark matter (DM) on hybrid star properties. We assume that dark matter is mixed with both hadronic and quark matter and interact with them through the exchange of a Higgs boson. The hybrid star properties are obtained from equations of state calculated with a Maxwell prescription. For the hadronic matter we use the NL3* parameter set and for the quark matter, the MIT bag model with a vector interaction. We see that dark matter does not influence the phase transition points (pressure and chemical potential) but shifts the discontinuity on the energy density, which ultimately reduces the minimum mass star that contains a quark core. Moreover, it changes considerably the star family mass-radius diagrams and moves the merger polarizability curves inside the confidence lines. Another interesting feature is the influence of DM in the quark core of the hybrid stars constructed. Our results show an increase of the core radius for higher values of the dark particle Fermi momentum.
\end{abstract}

\maketitle

\section{Introduction}

Cosmological and astrophysical data suggest that ordinary baryonic matter comprises only 5\% of the constituents of the Universe, the remainder being dark matter (approximately 23\%) and dark energy (approximately 72\%). Dark matter (DM) is called dark because it does not absorb, reflect or emit electromagnetic radiation and hence, it is very difficult to be detected, but it certainly feels the gravitational force. There are different candidates for dark matter, but its true nature and origin are a mystery. Among the candidates, some are of baryonic origin and others are non-baryonic. Strong candidates are the weakly interacting massive particles (WIMPS). For a review on the subject, the interested reader can look at \cite{bertone,pdg22}, among many other recent publications.

In recent years, many studies on the possibility that DM can be a part of compact objects, as neutron stars, and affect their macroscopic properties, as masses and radii, have been considered. An admixture of DM with hadronic matter has been extensively discussed in the literature \cite{rmfdm13,rmfdm1,rmfdm2,rmfdm3,abdul,rmfdm6,rmfdm11,rmfdm10,rmfdm8,rmfdm7,rmfdm12,dmnosso1,dmnosso2,dmnosso3,laura-tolos,sagun,Lopes_2018,karkevandi1,karkevandi2}. Along the same line, dark matter effects have been studied in quark stars~\cite{fraga}.

However, a recent study with a model-independent analysis based on the calculation of the sound velocity in different media suggests that quark cores are expected inside massive neutron stars~\cite{nature_2020}, as the ones detected in the last years \cite{Miller2021,Riley2021,romani}. The idea of hybrid stars containing a hadronic and a quark core is not a novelty and was first proposed in 1965 \cite{Ivanenko}. To build the equation of state (EoS) that describes hybrid stars, two constructions are commonly used: the simpler one considers that the hadronic and the quark
phases are in direct contact and just one of the two independent chemical potentials is continuous during the phase transition. It is commonly named Maxwell construction. The other prescription considers that both chemical potentials are continuous and a 
a mixed phase containing hadrons and deconfined quarks has to be constructed. For a discussion on the differences between stellar structures
obtained with both constructions and a review on the three above mentioned types of neutron stars (hadronic, quark and hybrid), please refer to \cite{debora-universe}.

In the present work, we revisit the idea of hybrid stars based on a Maxwell construction and include an admixture of dark matter. For the hadronic phase, we use a relativistic mean field model (RMF) within the NL3* parameter set \cite{snl3}, for the quark phase the MIT bag model with a vector interaction, as proposed in \cite{Lopes_2021} and DM is taken into account by considering the kinetic terms only for different Fermi momenta and the neutralino mass equal to 200 GeV. Details are given in the following section.

In the next sections, we introduce the formalism, present our results and conclusions and finish with useful remarks. 

\section{Formalism}\label{formalism}

In the next two subsections, we present the basic expressions used to describe first hadronic matter  and then quark matter
coupled to dark matter. We leave the construction of hydrib stars with a mixture of dark matter for the Results Section. 

\subsection{Hadronic model coupled to dark matter}

For the hadronic part of the hadron-quark model, we use the one described by the Lagrangian density given by
\begin{align}
&\mathcal{L}_{\mbox{\tiny HAD}} = \overline{\psi}(i\gamma^\mu\partial_\mu - M_{\mbox{\tiny nuc}})\psi 
+ g_\sigma\sigma\overline{\psi}\psi 
- g_\omega\overline{\psi}\gamma^\mu\omega_\mu\psi
\nonumber \\ 
&- \frac{g_\rho}{2}\overline{\psi}\gamma^\mu\vec{\rho}_\mu\vec{\tau}\psi
+\frac{1}{2}(\partial^\mu \sigma \partial_\mu \sigma - m^2_\sigma\sigma^2)
- \frac{A}{3}\sigma^3 - \frac{B}{4}\sigma^4 
\nonumber\\
&-\frac{1}{4}F^{\mu\nu}F_{\mu\nu} 
+ \frac{1}{2}m^2_\omega\omega_\mu\omega^\mu 
-\frac{1}{4}\vec{B}^{\mu\nu}\vec{B}_{\mu\nu} 
+ \frac{1}{2}m^2_\rho\vec{\rho}_\mu\vec{\rho}^\mu,
\label{dlag}
\end{align}
in which $\psi$, $\sigma$, $\omega^\mu$, and $\vec{\rho}_\mu$, represent, respectively, the nucleon and the exchanged mesons $\sigma$, $\omega$, and $\rho$. The masses of such particles are denoted by $M_{\mbox{\tiny nuc}}$, $m_\sigma$, $m_\omega$, and $m_\rho$, and the coupling constants are $g_\sigma$, $g_\omega$, $g_\rho$, $A$, and $B$. Furthermore, the tensors read $F_{\mu\nu}=\partial_\mu\omega_\nu-\partial_\nu\omega_\mu$ and $\vec{B}_{\mu\nu}=\partial_\mu\vec{\rho}_\nu-\partial_\nu\vec{\rho}_\mu$. 

The coupling with dark matter is done as in Refs.~\cite{dmnosso1,dmnosso2,dmnosso3}, i.e., we use the total Lagrangian density as follows
\begin{align}
\mathcal{L} &= \overline{\chi}(i\gamma^\mu\partial_\mu - M_\chi)\chi
+ \xi h\overline{\chi}\chi +\frac{1}{2}(\partial^\mu h \partial_\mu h - m^2_h h^2)
\nonumber\\
&+ f\frac{M_{\mbox{\tiny nuc}}}{v}h\overline{\psi}\psi + \mathcal{L}_{\mbox{\tiny HAD}},
\label{dlagtotal}
\end{align}
with the dark fermion given by the Dirac field $\chi$ with related mass $M_\chi$. In this approach, 
the interaction between $\chi$ and $\psi$ is due to the Higgs boson whose mass is $m_h=125$~GeV. The strength of this interaction is controlled by the constant $fM_{\mbox{\tiny nuc}}/v$, where $v=246$~GeV is the Higgs vacuum expectation value. The constant $\xi$ is the Higgs-dark particle coupling. The mean-field approximation is used to compute the field equations, in this case given by
\begin{align}
m^2_\sigma\,\sigma &= g_\sigma\rho_s - A\sigma^2 - B\sigma^3 
\\
m_\omega^2\,\omega_0 &= g_\omega\rho, 
\\
m_\rho^2\,\bar{\rho}_{0(3)} &= \frac{g_\rho}{2}\rho_3, 
\\
[\gamma^\mu (&i\partial_\mu - g_\omega\omega_0 - g_\rho\bar{\rho}_{0(3)}\tau_3/2) - M^*]\psi = 0,
\\
m^2_h\,h &= \xi\rho_s^{\mbox{\tiny DM}} + f\frac{M_{\mbox{\tiny nuc}}}{v}\rho_s
\\
(\gamma^\mu &i\partial_\mu - M_\chi^*)\chi = 0,
\end{align}
with $\tau_3=1$ for protons and $-1$ for neutrons. In Ref.~\cite{dmnosso1}, the authors investigated the influence of $f$ and $\xi$ by using the values of $0.001\leqslant\xi\leqslant 0.1$~\cite{ilidio} and $f=0.3\pm0.015$~\cite{cline,cline-errata} and concluded that these quantities do not play any significant role, as discussed later in the present text. 

The effective nucleon and dark particle masses are 
\begin{eqnarray}
M^* = M_{\mbox{\tiny nuc}} - g_\sigma\sigma - f\frac{M_{\mbox{\tiny nuc}}}{v}h
\end{eqnarray}
and 
\begin{eqnarray}
M^*_\chi = M_\chi - \xi h,
\end{eqnarray}
respectively, and the densities are $\rho_s=\left<\overline{\psi}\psi\right>={\rho_s}_p+{\rho_s}_n$, $\rho=\left<\overline{\psi}\gamma^0\psi\right> = \rho_p + \rho_n$, $\rho_3=\left<\overline{\psi}\gamma^0{\tau}_3\psi\right> = \rho_p - \rho_n=(2y_p-1)\rho$, and
$\rho_s^{\mbox{\tiny DM}} = \left<\overline{\chi}\chi\right>$, with
\begin{eqnarray}
\rho_s^{\mbox{\tiny DM}} &=& 
\frac{\gamma M^*_\chi}{2\pi^2}\int_0^{k_F^{\mbox{\tiny DM}}} \hspace{-0.5cm}\frac{k^2dk}{(k^2+M^{*2}_\chi)^{1/2}}.
\end{eqnarray}
$\gamma=2$ is the degeneracy factor, and the proton fraction is denoted by $y_p=\rho_p/\rho$, with $\rho_{p,n}=\gamma{k_F^3}_{p,n}/(6\pi^2)$. The Fermi momenta related to protons, neutrons, and dark particle are respectively ${k_F}_{p,n}$ and $k_F^{\mbox{\tiny DM}}$. 

The main thermodynamical quantities related to this hadron-DM system, namely, energy density and pressure, are obtained from the energy-momentum tensor $T^{\mu\nu}$. In our case, Eq.~(\ref{dlagtotal}) is used to calculate $\mathcal{E}_{\mbox{\tiny HAD-DM}}=\left<T_{00}\right>$ and $P_{\mbox{\tiny HAD-DM}}=\left<T_{ii}\right>/3$. This procedure gives rise to the following expressions,
\begin{align} 
&\mathcal{E}_{\mbox{\tiny HAD-DM}} = \frac{m_{\sigma}^{2} \sigma^{2}}{2} +\frac{A\sigma^{3}}{3} +\frac{B\sigma^{4}}{4} 
-\frac{m_{\omega}^{2} \omega_{0}^{2}}{2} - \frac{m_{\rho}^{2} \bar{\rho}_{0(3)}^{2}}{2} 
\nonumber\\
&+ g_{\omega} \omega_{0} \rho + \frac{g_{\rho}}{2} 
\bar{\rho}_{0(3)} \rho_{3} + \frac{m_h^2h^2}{2} + \mathcal{E}_{\mathrm{kin}}^{p} + \mathcal{E}_{\mathrm{kin}}^{n} + \mathcal{E}_{\mathrm{kin}}^{\mbox{\tiny DM}},
\label{eden}
\end{align}
and
\begin{align}
&P_{\mbox{\tiny HAD-DM}} = -\frac{m_{\sigma}^{2} \sigma^{2}}{2} - \frac{A\sigma^{3}}{3} - \frac{B\sigma^{4}}{4} + \frac{m_{\omega}^{2} \omega_{0}^{2}}{2} + \frac{m_{\rho}^{2} \bar{\rho}_{0(3)}^{2}}{2}
\nonumber\\
& - \frac{m_h^2h^2}{2} + P_{\mathrm{kin}}^{p} + P_{\mathrm{kin}}^{n} + P_{\mathrm{kin}}^{\mbox{\tiny DM}},
\label{press}
\end{align}
with the kinetic terms related to the dark particle given by
\begin{eqnarray}
\mathcal{E}_{\mathrm{kin}}^{\mbox{\tiny DM}} &=& \frac{\gamma}{2\pi^2}\int_0^{k_F^{\mbox{\tiny DM}}}\hspace{-0.3cm}k^2(k^2+M^{*2}_\chi)^{1/2}dk,
\label{ekindm}
\end{eqnarray}
and
\begin{eqnarray}
P_{\mathrm{kin}}^{\mbox{\tiny DM}} &=& 
\frac{\gamma}{6\pi^2}\int_0^{{k_F^{\mbox{\tiny DM}}}}\hspace{-0.5cm}\frac{k^4dk}{(k^2+M^{*2}_\chi)^{1/2}}.
\label{pkindm}
\end{eqnarray}
Proton and neutron kinetic terms are defined as in Eqs.~(\ref{ekindm})-(\ref{pkindm}) by taking into account the following replacements: $k_F^{\mbox{\tiny DM}}\rightarrow {k_F}_{p,n}$ and $M^*_\chi\rightarrow M^*$.

As shown in Ref.~\cite{dmnosso1}, the formulation used to couple dark matter to the hadronic one leads to the modification in the energy density and pressure only by adding the dark particle kinetic terms. Therefore, it is totally safe to rewrite Eqs.~(\ref{eden})-(\ref{press}) as 
\begin{align} 
&\mathcal{E}_{\mbox{\tiny HAD-DM}} = \frac{m_{\sigma}^{2} \sigma^{2}}{2} +\frac{A\sigma^{3}}{3} +\frac{B\sigma^{4}}{4} 
-\frac{m_{\omega}^{2} \omega_{0}^{2}}{2} - \frac{m_{\rho}^{2} \bar{\rho}_{0(3)}^{2}}{2} 
\nonumber\\
&+ g_{\omega} \omega_{0} \rho + \frac{g_{\rho}}{2} 
\bar{\rho}_{0(3)} \rho_{3} + \mathcal{E}_{\mathrm{kin}}^{p} + \mathcal{E}_{\mathrm{kin}}^{n} + \mathcal{E}_{\mathrm{kin}}^{\mbox{\tiny DM}},
\label{edeneff}
\end{align}
and
\begin{align}
&P_{\mbox{\tiny HAD-DM}} = -\frac{m_{\sigma}^{2} \sigma^{2}}{2} - \frac{A\sigma^{3}}{3} - \frac{B\sigma^{4}}{4} + \frac{m_{\omega}^{2} \omega_{0}^{2}}{2} + \frac{m_{\rho}^{2} \bar{\rho}_{0(3)}^{2}}{2}
\nonumber\\
& + P_{\mathrm{kin}}^{p} + P_{\mathrm{kin}}^{n} + P_{\mathrm{kin}}^{\mbox{\tiny DM}},
\label{presseff}
\end{align}
due to the smallness of the scalar field $h$. In this case, the nucleon effective mass takes its traditional form $M^* = M_{\mbox{\tiny nuc}} - g_\sigma\sigma$, and the dark particle remains constant, i.e., $M^*_\chi = M_\chi=200$~GeV (we also use here the lightest neutralino as the dark particle candidate~\cite{cand1,cand2}). We use the Fermi momentum to fix the DM content by using different values for this quantity, namely, $k_F^{\mbox{\tiny DM}}=0$ (no DM included), $0.02$~GeV, $0.04$~GeV, and $0.06$~GeV. This implies in constant contributions to the energy density and pressure as well.

As a last remark of this section, we emphasize that the parametrization used in the hadronic part of the system is the NL3* parameter set~\cite{snl3}. It was recently selected in a systematic study in which finite nuclei properties were analyzed, as well as neutron stars ones
\cite{brett-jerome}. This particular parametrization, among other ones, reproduce experimental data of ground state binding energies, charge radii and giant monopole resonances of a set of spherical nuclei and it is also in agreement with some stellar matter constraints.

\subsection{Effective quark model: vector MIT bag model coupled to dark matter}

We use the thermodynamic consistent vector MIT bag model introduced in Ref.~\cite{Lopes_2021, Lopes_2022} to describe the quark matter. In this model, the quark interaction is mediated by the vector channel $V^\mu$, analogous to the $\omega$ meson in QHD~\cite{Serot_1992}. Indeed, in this work, we consider that the vector channel is the $\omega$ meson itself. Its Lagrangian density reads:
%%%%%%%%%%%%%%%%
\begin{eqnarray}
\mathcal{L}_{\rm vMIT} &=& \bigg\{ \bar{\psi}_q\big[\gamma^\mu(i\partial_\mu - g_{qV} V_\mu) - m_q\big]\psi_q  \nonumber \\
&&
- B + \frac{1}{2}m_V^2V^\mu V_\mu  \bigg\}\Theta(\bar{\psi}_q\psi_q) ,
\label{vMIT}
\end{eqnarray}
%%%%%%%%%%%%%%%
where $m_q$ is the mass of the quark $q$ of ﬂavor $u$, $d$ or $s$, $\psi_q$ is the Dirac quark ﬁeld, $B$ is the constant vacuum pressure, and $\Theta(\bar{\psi}_q\psi_q)$ is the Heaviside step function to assure that the quarks exist only conﬁned to the bag. Applying the Euler-Lagrange equations, we obtain the energy eigenvalue, which at $T = 0$ K, is also the chemical potential:
%%%%%%%%%%%%%%%%
\begin{equation}
E_q = \mu_q = \sqrt{m_q^2 + k^2} + g_{qV}V_\mu,
\label{E}
\end{equation}
%%%%%%%%%%%%%%%
now, using Fermi-Dirac statistics, we are able to obtain the EoS in mean field approximation. The energy density of the quarks is:
%%%%%%%%%%%%%%%%
\begin{equation}
\epsilon_q = \frac{N_c}{\pi^2}\int_0^{k_{F_q}} E_q k^2 d^3k , \label{ed}
\end{equation}
%%%%%%%%%%%%%%%
where $N_c = 3$ is the number of colors and $k_{F_q}$ is the Fermi momentum of the quark $q$. The contribution of the bag, as well as the mesonic mass term, is obtained with the Hamiltonian: $\mathcal{H}$ = $- \langle \mathcal{L} \rangle$. The total quark energy density now reads:
%%%%%%%%%%%%%%%%
\begin{equation}
\mathcal{E}_{\mbox{\tiny quarks}} = \sum_q\epsilon_q + B - \frac{1}{2}m_v^2V_0^2. \label{ted}
\end{equation}
%%%%%%%%%%%%%%%
% In order to construct an electric neutral, beta stable matter, leptons are added as a free Fermi gas. 
The pressure is obtained via the relation
\begin{eqnarray}
P_{\mbox{\tiny quarks}} = \sum_q \mu_q n_q - \mathcal{E}_{\mbox{\tiny quarks}}
\end{eqnarray}
where the sum runs over all the quarks.

The parameters utilized in this work are the same as presented in Ref.~\cite{Lopes_2021}. We use $m_u = m_d$ = 4 MeV, and $m_s$ = 95 MeV. We also assume an universal coupling of quarks with the vector meson, i.e, $g_{uV} = g_{dV} = g_{sV} = g_V$, and use some  values of $G_V$; as defined below
\begin{equation}
G_V = \bigg ( \frac{g_V}{m_V} \bigg )^2,
\label{GV}
\end{equation}
in units of fm$^{2}$. The value of the bag is taken as $B^{1/4}$ = 158 MeV. The coupling of quark matter to dark matter is done by considering the result presented in the last section, namely, the modification in the equations of state
%the quark ones in this case, 
is only due to the inclusion of the DM kinetic terms in the energy density and pressure, what leads to
\begin{equation}
\mathcal{E}_{\mbox{\tiny Q-DM}} = \mathcal{E}_{\mbox{\tiny quarks}}
+ \mathcal{E}_{\mathrm{kin}}^{\mbox{\tiny DM}} 
\label{edquarksdm}
\end{equation}
and
\begin{eqnarray}
P_{\mbox{\tiny Q-DM}} = P_{\mbox{\tiny quarks}} + P_{\mathrm{kin}}^{\mbox{\tiny DM}} 
\label{pressquarksdm}
\end{eqnarray}

% \subsection{The Fermionic Dark matter model}

% The Lagrangian of the fermionic DM reads:
% %%%%%%%%%%%%%%%%
% \begin{eqnarray}
% \mathcal{L}_{DM} &=& \bar{\chi}(i \gamma^\mu \partial_\mu - (m_x -g_H h))\chi 
% \nonumber \\
% &&
% + \frac{1}{2}(\partial^\mu h \partial_\mu h - m_H^2 h^2). \label{FDMEOS}
% \end{eqnarray}
% %%%%%%%%%%%%%%%
% Here, we assume a dark fermion represented by the Dirac field $\chi$ that self-interacts through the exchange of the Higgs boson, whose mass is $m_H$ = 125 GeV. The coupling constant is assumed to be $g_H = 0.01$,  which agrees with the constraints in Refs.~\cite{Panotopoulos_2017, Das_2021}. Within this prescription, the DM self-interaction is very feeble and behaves as a free Fermi gas. More explicitly:
% %%%%%%%%%%%%%%%%
% \begin{equation}
% G_H = \bigg ( \frac{g_H}{m_H} \bigg )^2 = 2.492 \times 10^{-10} \quad \mbox{fm}^{2} .
% \end{equation}
% %%%%%%%%%%%%%%%
% The EoS is easily obtained in mean field approximation, completely analogous to the QHD model~\cite{Serot_1992}. The fermionic DM is assumed to be the lightest neutralino, with $m_x$ = 200 GeV, as done in Ref.~\cite{OdilonDM, Das_2021}.

\section{Results} 

Next, we first explain how the hybrid star equation of state is built and then discuss its main astrophysical properties. 

\subsection{Hybrid Star Equation of States}

In this work we assume that the phase transition between hadrons and quarks is described by the Maxwell construction, thereby the pressure and baryonic chemical potential  have the same value at the interface. The hybrid EoSs are constructed combining some different values of the dark matter contribution, $k_F^{\mbox{\tiny DM}}$ associated with different vector coupling constant, $G_V$. The hadronic and quark models are described in Sec. \ref{formalism}. 
As already pointed out in previous works, the larger the dark matter Fermi momentum, the softer, the resulting EoS~\cite{rmfdm2,rmfdm3,abdul,rmfdm6,rmfdm8,rmfdm10,rmfdm11}.

\begin{figure}
%\begin{center}
\includegraphics[trim=0cm 6cm 0cm 6cm, clip, scale=0.43]{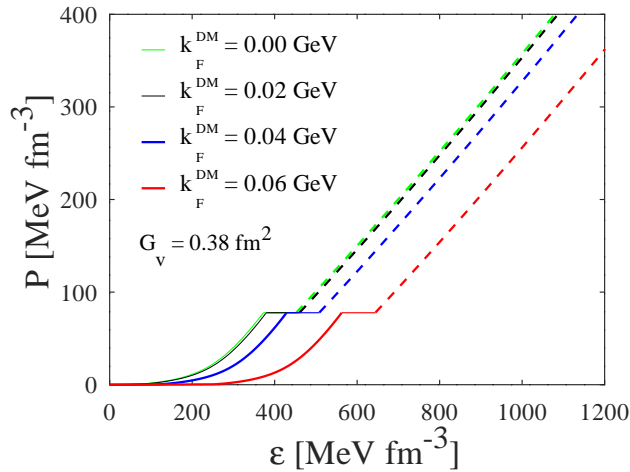}
\caption{Pressure as a function of energy density of hybrid EoSs for different values of $k_F^{\mbox{\tiny DM}}$ using one fixed value for the vector coupling constant $G_V$. Full and dashed lines indicate hadronic and quark sectors, respectively.}
\label{pt}
%\end{center}
\end{figure}

Fig. \ref{pt} shows the hybrid EoSs for different values of $k_F^{\mbox{\tiny DM}}$ when $G_V = 0.38 ~\mbox {fm}^{2}$.
 The ``{\it gaps}'' in the energy density are due to the Maxwell construction. It is possible to note that in our formalism, the dark matter contribution has no effect on the transition points defined by the Gibbs conditions, $P_t$ and $\mu_t$.
 %, the amplitude of the gaps $\Delta \varepsilon$. 
 On the other hand the starting value, $\epsilon_i$, of the phase transition is very sensitive to $k_F^{\mbox{\tiny DM}}$. In Table \ref{table2} we %can see as we can control 
display %the parameters of 
some of the values associated with the first order phase transition, namely, $P_t$, $\Delta \epsilon$, the chemical potential, $\mu_t$, and $\epsilon_i$ for different dark matter momentum, $k_F^{\mbox{\tiny DM}}$ and six values of the vector coupling constant, $G_V$.
One well known feature is that the transition point moves to higher chemical potentials and pressures when $G_V$ increases \cite{Lopes_2022} and this fact can be easily observed in Table \ref{table2}.
\begin{table}
\setlength{\tabcolsep}{0.5pt}
\renewcommand{\arraystretch}{1.3}
\caption{Chemical potential ($\mu_t$), pressure ($P_t$), energy density gap ($\Delta \epsilon$) and initial value where the phase transition begins, $\epsilon_i$, related to the transition for different coupling constants of the vector MTI bag model, $G_V$, and dark matter Fermi momentum, $k_F^{\mbox{\tiny DM}}$. Note that the dark matter contribution is relevant only for the initial energy density, $\epsilon_i$.}
\begin{center}
\begin{tabular}{|c|c|c|c|c|c|c|c|} \hline
$G_V$              &   $\mu_t$   &    $P_t$ & $\Delta \epsilon$ & \multicolumn{4}{c|}{$k_F^{\mbox{\tiny DM}}$~(GeV)}  \\ \cline{5-8}
$(\mbox{fm}^{2})$ &  (MeV)      &   $(\mbox{MeV}/\mbox{fm}^3)$ & $(\mbox{MeV}/\mbox{fm}^3)$ & 0.00 & 0.02 & 0.04 & 0.06 \\ \cline{5-8} 
\ \                &    \ \      &  \ \   &  \ \ &  \multicolumn{4}{c|}{$\epsilon_i~(\mbox{MeV.fm}^{-3})$} \\ \hline\hline
0.20 & 1004.75  & 7.12   & 137.0 & 170.4 & 175.9 & 225.2 & 358.8 \\ 
0.32 & 1128.79  & 37.28  & 84.4  & 292.4  &  297.8  & 348.7  & 482.4  \\
0.35 & 1184.25  & 54.14  & 78.0  & 328.8  & 335.9  & 385.1   & 518.7 \\
0.38 & 1252.91  & 77.93  & 79.8  & 372.6  & 379.6  & 428.8  & 562.4  \\
0.50 & 1649.65  & 256.84 & 139.5 & 630.6 & 637.6 & 686.9  & 820.5 \\
0.60 & 2141.42  & 574.19 & 245.3 & 1036.9 & 1044.0 & 1093.2  & 1226.8 \\ \hline
\end{tabular}
\end{center}
\label{table2}
\end{table}

\subsection{The mass-radius relation}

To construct hydrostatic stellar configurations, we use the Tolman-Oppnheimer-Volkoff (TOV) equations~\cite{tov39,tov39a} given by
\begin{align}
\frac{\mathrm{d} P}{\mathrm{~d} r} & =-\frac{m(r)  \epsilon(r)}{r^{2}} \frac{[1+P(r)  / \epsilon(r) ]\left[1+4 \pi r^{3} P(r)  / m(r) \right]}{g(r)} ,  \\
\frac{\mathrm{d} m}{\mathrm{~d} r} & =4 \pi r^{2} \epsilon  , 
\end{align}
where $g(r)=1-2m(r)/r$, $m(r)$ is the gravitational mass enclosed within the radial coordinate $r$, $P(r)$ and $\epsilon(r)$ are the pressure and energy density at a $r$, and we consider $G=c=1$. To solve the TOV equations we 
%have to require the
need boundary conditions at the stellar center and at the surface of the star and we take 
$m(r=0)=0$ and $P(r=R)=0$, respectively.

An important aspect to be considered when analysing the hybrid star stability 
%in hybrid stars scenarios 
is its response to small radial perturbations, see for instance Refs.~\cite{rad1,rad2,rad3,rad4,rad5,rad6,rad7} for details. According to radial oscillations theory, stars are considered stable when the oscillations due to radial perturbations are well defined. On the other hand, if the radial perturbations amplitude presents an indefinite increase, an unstable star is characterized.  

 Hence, when we analyse radial oscillations in a hybrid star scenario we have to take into account the kind of phase transition
 that takes place.
In stars with two different matter constitutions, the phase transition can be classified as ``{\it fast}'' or ``{\it slow}''~\cite{rad0}. The slow transitions occur when the timescale of the reactions near the interface is greater than radial oscillations. As a consequence, the fluids on both sides of the interface maintain their compositions and co-move with the interface of phase transition. Such implications are encoded in the junction conditions \cite{rad0} and the consequences can be seen in mass-radius diagram, where it is possible to note that the region of stable star can be found beyond the maximum mass point~\cite{rad3,rad4,rad2}.  On the other hand, in the case of a fast phase transition, the timescale of reactions transforming one phase into another near the interface is lower than the one associated with radial perturbations. Thus, in this scenario the mass flow through the interface occurs and, as a consequence, all stars beyond maximum mass are unstable~\cite{rad2}.
 
\begin{figure*}
%\begin{center}
\includegraphics[trim=0cm 6cm 0cm 6cm, clip, scale=0.35]{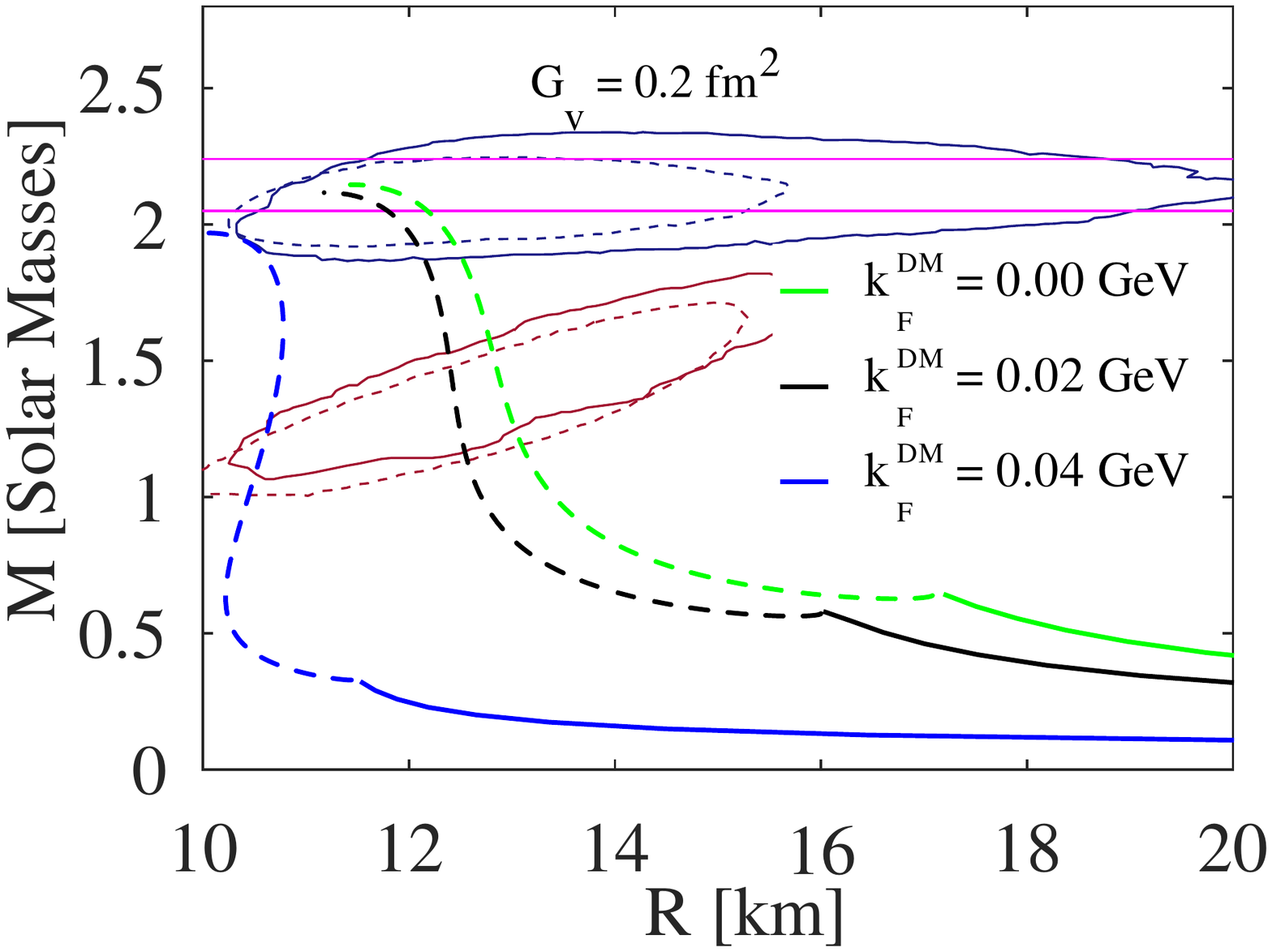} 
\includegraphics[trim=0cm 6cm 0cm 6cm, clip, scale=0.35]{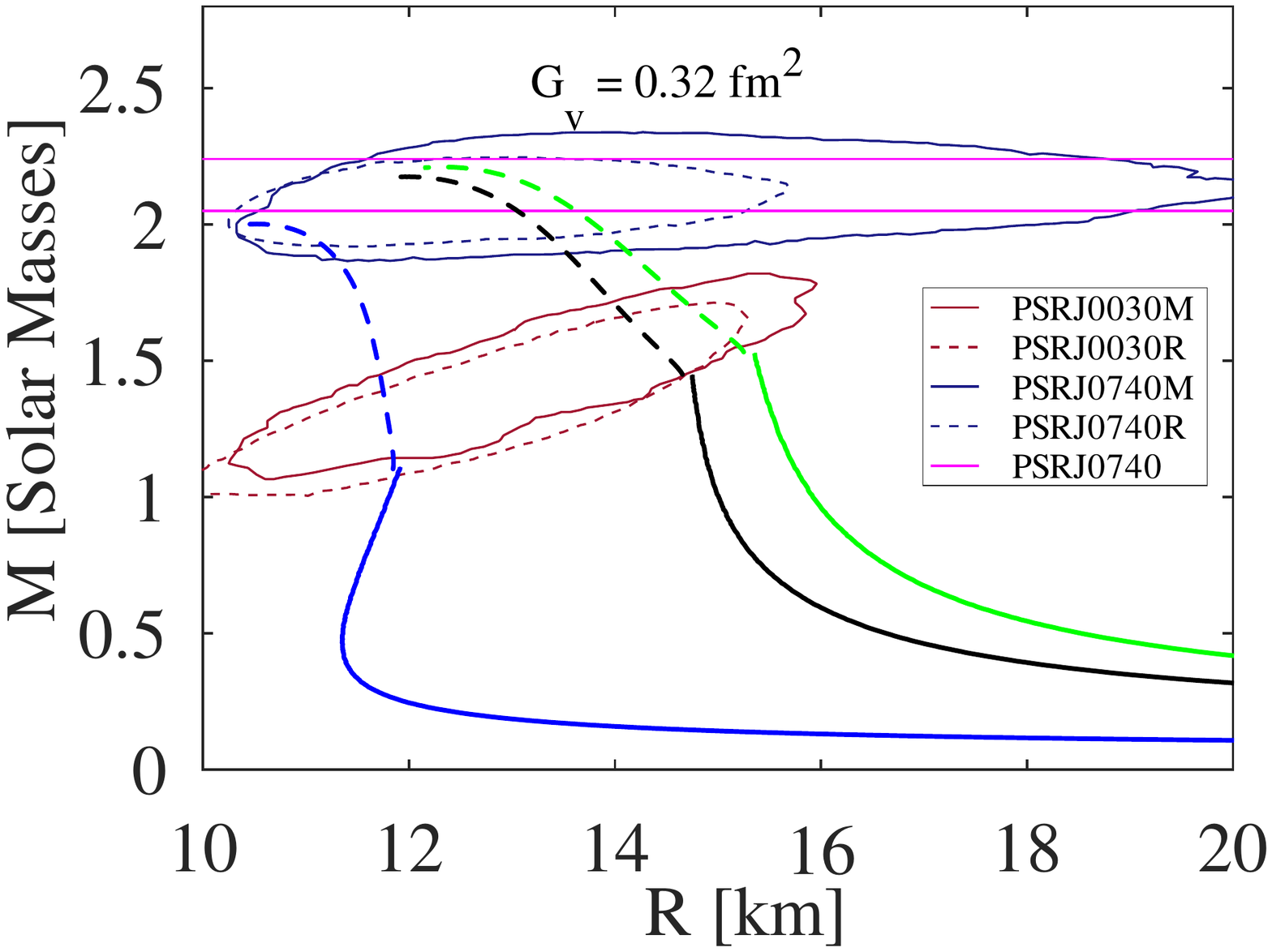} \\
\includegraphics[trim=0cm 6cm 0cm 6cm, clip, scale=0.35]{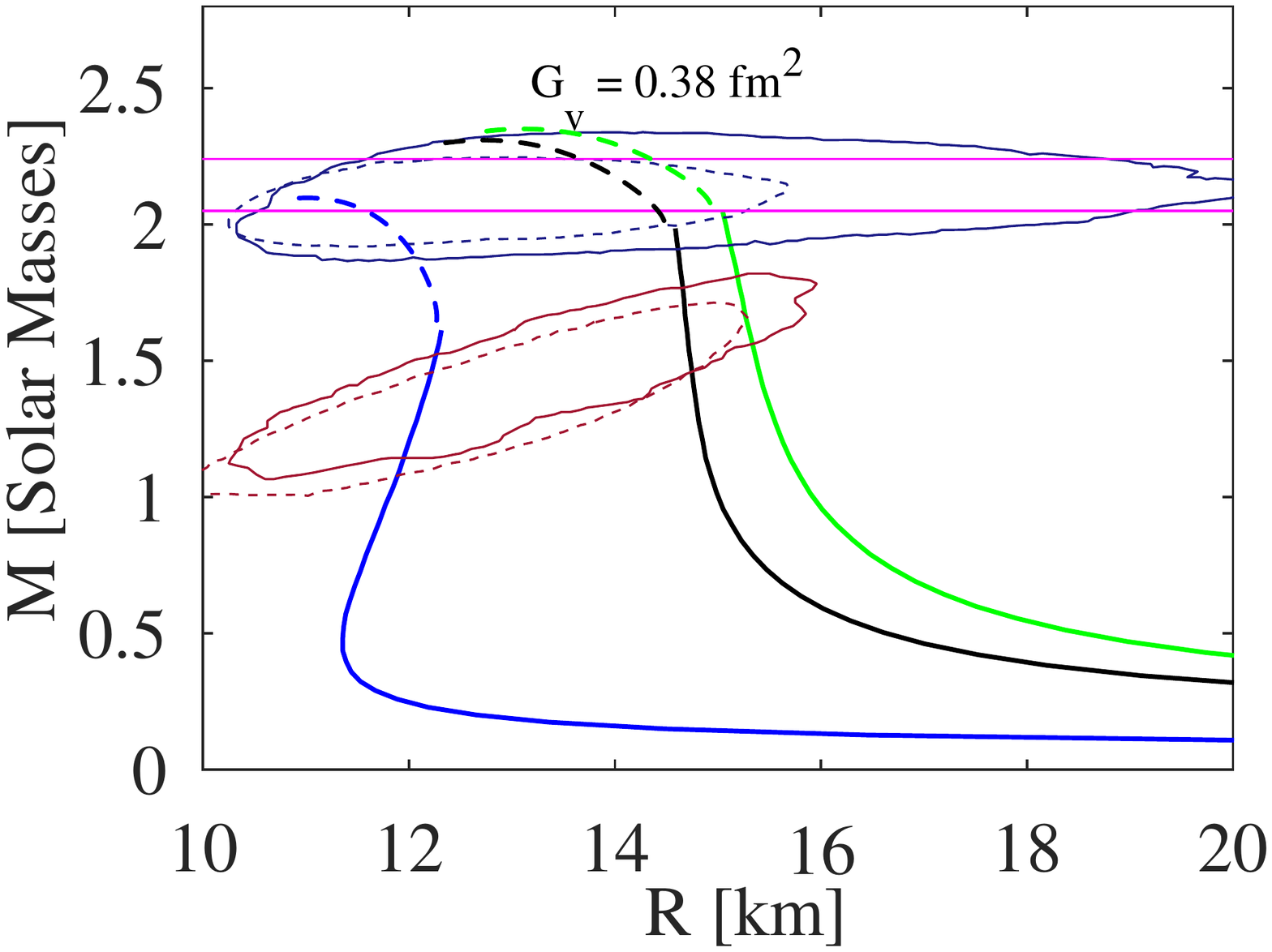} 
\includegraphics[trim=0cm 6cm 0cm 6cm, clip, scale=0.35]{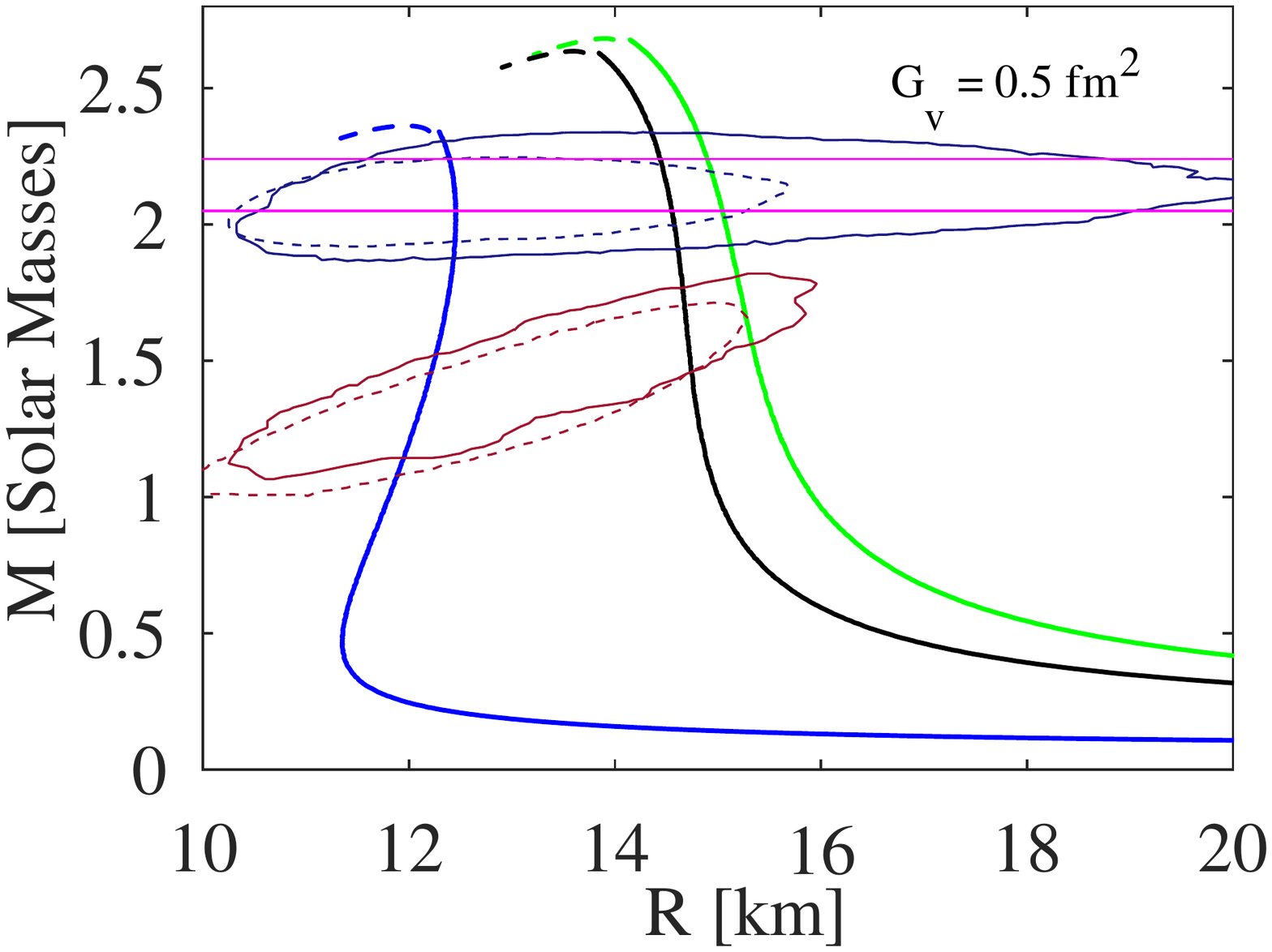} 
\caption{Mass as a function of the radii for the hybrid EoSs for different values of $k_F^{\mbox{\tiny DM}}$ and $G_V$. Full and dashed lines indicate hadronic and hybrid sector in the sequences, respectively. In all diagrams are shown constraints from the $\sim 2~M_\odot$ pulsars and NICER observations.}
\label{MR}
%\end{center}
\end{figure*}

The mass-radius diagram for the hybrid EoSs shown in Fig.~\ref{pt} can be seen in Fig.~\ref{MR}. 
In all cases we compute our results until the last stable star in the light of radial perturbation taking into account slow phase transitions. Analysing our results it is clear that the increase of the dark matter contribution decreases the radius and maximum mass in all cases of $G_V$. Furthermore, it is possible to realise that $k_F^{\mbox{\tiny DM}}$ favors the emergence of larger %and larger 
quark cores. Note also that all models in Fig.~\ref{MR} are in agreement with  constraints from the $2M_\odot$ pulsars and NICER observations \cite{Riley:2019yda,Miller:2019cac,Miller2021,Riley2021}. In particular, when $G_V = (0.2, 0.32) ~\mbox{fm}^{2}$ we can see that the hybrid star sectors (dashed lines) are predominant on the observable regions.

\begin{figure}
\begin{center}
\includegraphics[trim=0cm 6cm 0cm 6cm, clip, scale=0.4]{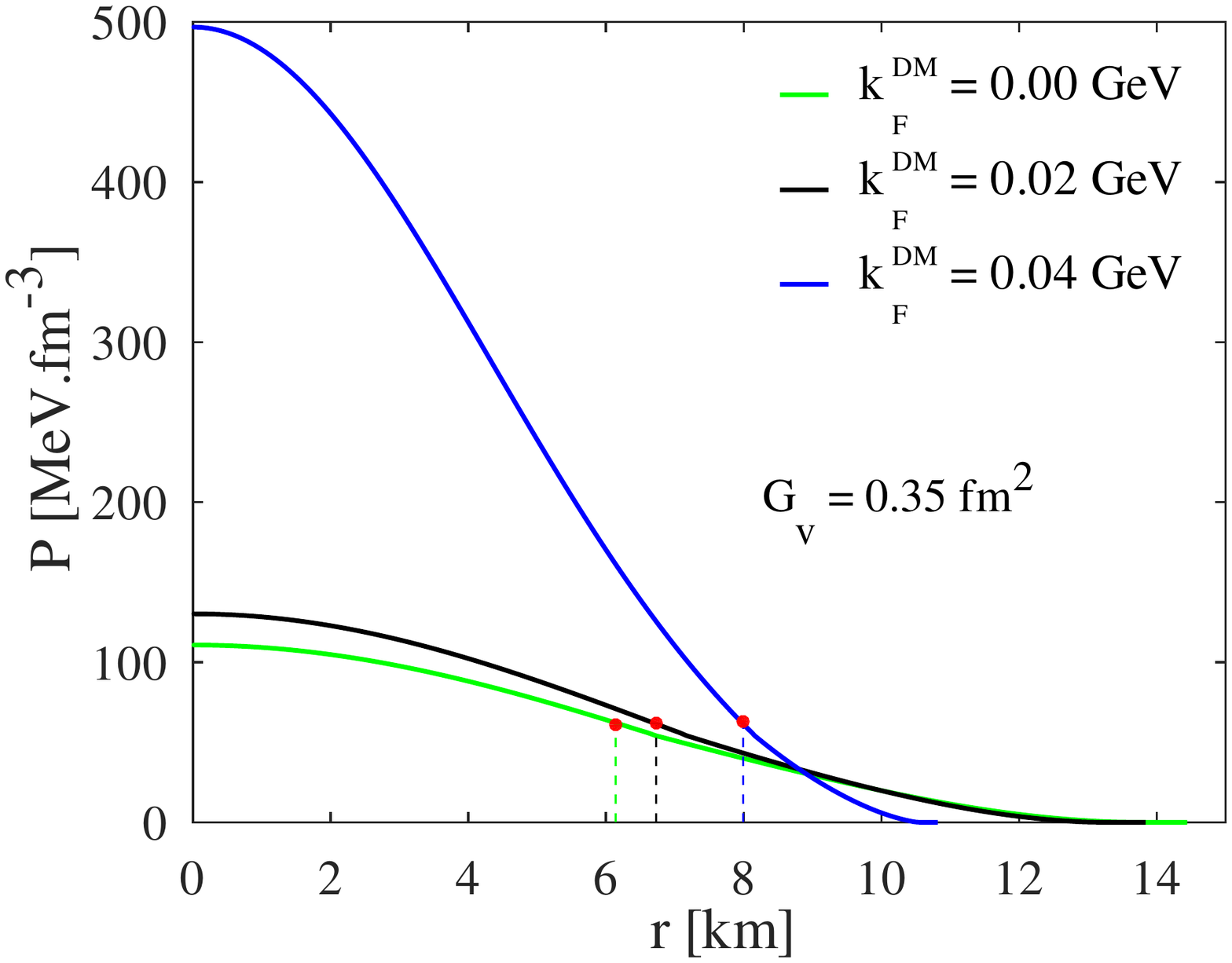}
\caption{Pressure as a function of the radial coordinate of a star with $2M_\odot$ for different contributions of dark matter. The red dots mark the radial coordinate point where occurs the phase transition between quark and hadronic phases.}
\label{perfil}
\end{center}
\end{figure}

In Fig.~\ref{perfil} we show the pressure as a function of the radial coordinate of a star with $2M_{\odot}$, for 
%the case when the vector coupling constant is
$G_V = 0.38~\mbox{fm}^{2}$ and three different values of $k_F^{\mbox{\tiny DM}}$. 
%In the same diagram w
We mark the radial point where the interface between the hadronic and  the
quark phases takes place. As we can see in Fig.~\ref{MR}, the radius of the star decreases with increasing values of $k_F^{\mbox{\tiny DM}}$. On the other hand, in Fig.~\ref{perfil} is clear that quark core increases with the increase of dark matter momentum. In this sense, the results shown in  Table~\ref{table3} confirm that %the percentage of 
the quark core size compared with the maximum radii of the star increases with $k_F^{\mbox{\tiny DM}}$.       
\begin{table}
\setlength{\tabcolsep}{10pt}
\renewcommand{\arraystretch}{1.3}
\caption{Quark core size, $r_{core}$, the maximum radii and the fraction, $r_{core}/R$ for different coupling constants of the vector MTI bag model, $G_V$, and dark matter Fermi momentum, $k_F^{\mbox{\tiny DM}}$. Note that the dark matter contribution increases the core size.}
\begin{center}
\begin{tabular}{|c|c|c|c|c|} \hline
$G_V$ & $k_F^{\mbox{\tiny DM}}$ & $r_{core}$ & $R$ & $r_{core}/R$ \\
(fm$^{2}$) & (GeV) & (km) & (km) &   (\%) \\ \hline \hline
0.32  & 0.00 &  8.24  & 13.80 &  59.7 \\ \hline
0.32  & 0.02 &  8.44  & 13.23 &  63.8  \\ \hline
0.32  & 0.04 &  8.48  & 10.62 &  79.8 \\ \hline \hline
0.35  & 0.00 &  6.14  & 14.54 &  42.2 \\ \hline
0.35  & 0.02 &  6.73  & 13.92 &  48.3 \\ \hline
0.35  & 0.04 &  7.99  & 11.29 &  70.9 \\  \hline
\end{tabular}
\end{center}
\label{table3}
\end{table}

\subsection{Tidal Deformability Parameter}

Another important astrophysical constraint comes from the GW170817 event, detected by the LIGO/Virgo gravitational wave telescopes: the dimensionless tidal deformability parameter $\Lambda$.
The tidal deformability of a compact object is a single parameter that quantiﬁes how easily the object is deformed when subjected to an external gravitational ﬁeld. Larger tidal deformability indicates that the object is easily deformable. On the opposite side, a compact object with a smaller tidal deformability parameter is smaller, more compact, and it is more difﬁcult to deform. It is deﬁned as
\begin{equation}
\Lambda = \frac{2k_2}{3C^5}, \label{tidal}
\end{equation}
where $C = M/R$ is the compactness of the star. The parameter $k_2$ is called the Love number and is related to the metric perturbation. It is given by
\begin{eqnarray}
&k_2& = \frac{8C^5}{5}(1-2C)^2[2+2C(y_R-1)-y_R]\nonumber\\
&\times&\Big\{2C [6-3y_R+3C(5y_R-8)] \nonumber\\
&+& 4C^3[13-11y_R+C(3y_R-2) + 2C^2(1+y_R)]\nonumber\\
&+& 3(1-2C)^2[2-y_R+2C(y_R-1)]{\rm ln}(1-2C)\Big\}^{-1},\qquad
\label{k2}
\end{eqnarray}
with $y_R\equiv y(R)$. The function $y(r)$ is obtained through the solution of $r(dy/dr) + y^2 + yF(r) + r^2Q(r)=0$, solved together with TOV equations. The quantities $F(r)$ and $Q(r)$ read
\begin{eqnarray}
F(r) &=& \frac{1 - 4\pi r^2[\epsilon(r) - P(r)]}{g(r)} , 
\\
Q(r)&=&\frac{4\pi}{g(r)}\left[5\epsilon(r) + 9P(r) + 
\frac{\epsilon(r)+P(r)}{v_s^2(r)}- \frac{6}{4\pi r^2}\right]
\nonumber\\ 
&-& 4\left[ \frac{m(r)+4\pi r^3 P(r)}{r^2g(r)} \right]^2,
\label{qr}
\end{eqnarray}
respectively, where the squared sound velocity is  $v_s^2(r)=\partial P(r)/\partial\epsilon(r)$. We address Refs.~\cite{rad2,Abbott:2018wiz,ferrer,chatziioannou,Flores2020}, and references therein, to the interested reader for a more complete discussion about the Love number and its calculation procedure. Regarding the calculation of $y(r)$ for hybrid stars~\cite{rad2}, we emphasize that this quantity presents a singularity due to the energy density discontinuity. In order to avoid this problem, the following junction conditions must be imposed
\begin{equation}
y(r_d + \epsilon') = y(r_d - \epsilon') - \frac{\Delta \epsilon}{m(r_d)/(4\pi r_d^3) + P(r_d)},
\end{equation}
where $r_d$ represents the point inside the star where the phase transition occurs. More details about the junction condition can be find in \cite{rad2,takatsy} and references therein.

\begin{figure*}
%\begin{center}
\includegraphics[trim=0cm 6cm 0cm 6cm, clip, scale=0.33]{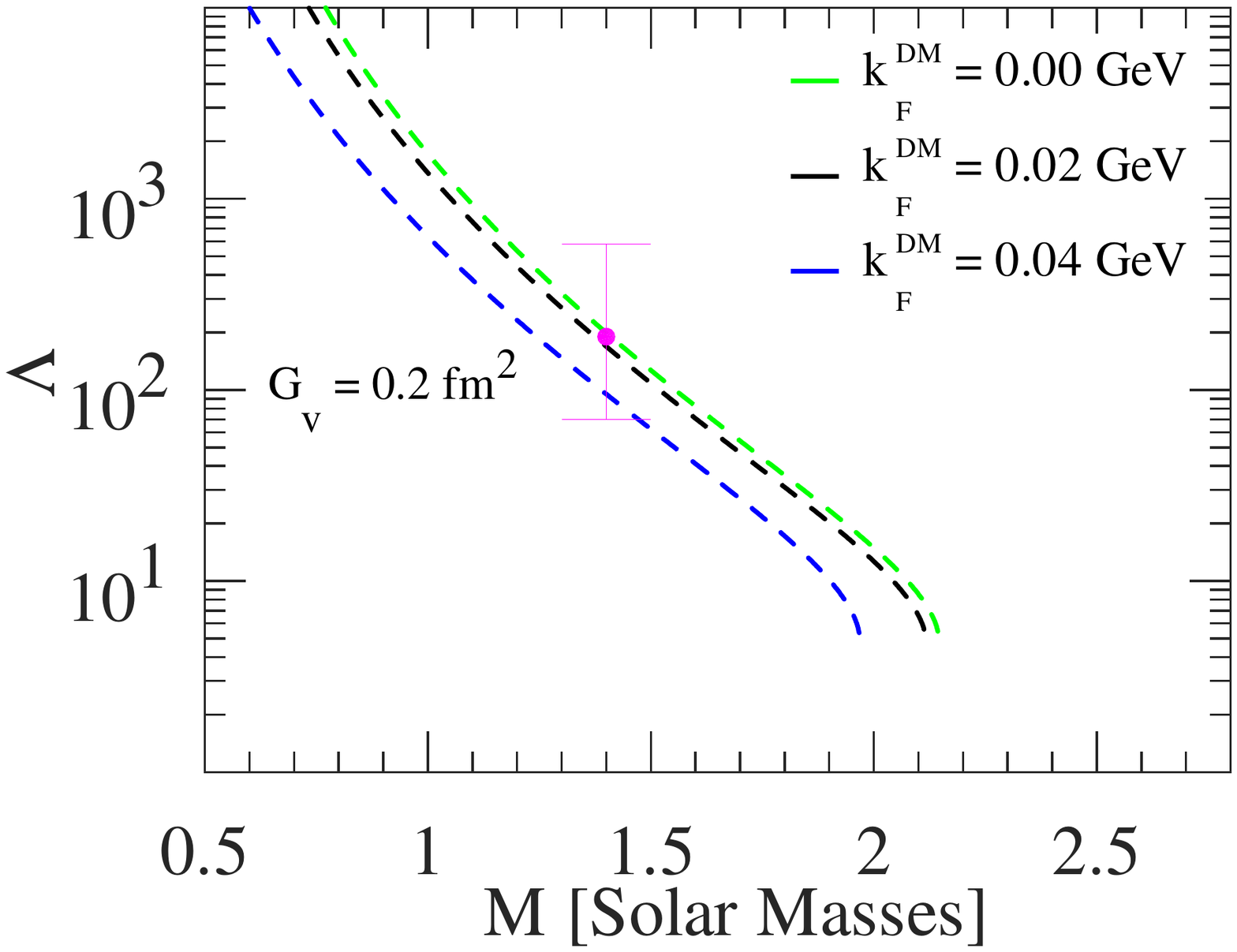}
\includegraphics[trim=0cm 6cm 0cm 6cm, clip, scale=0.33]{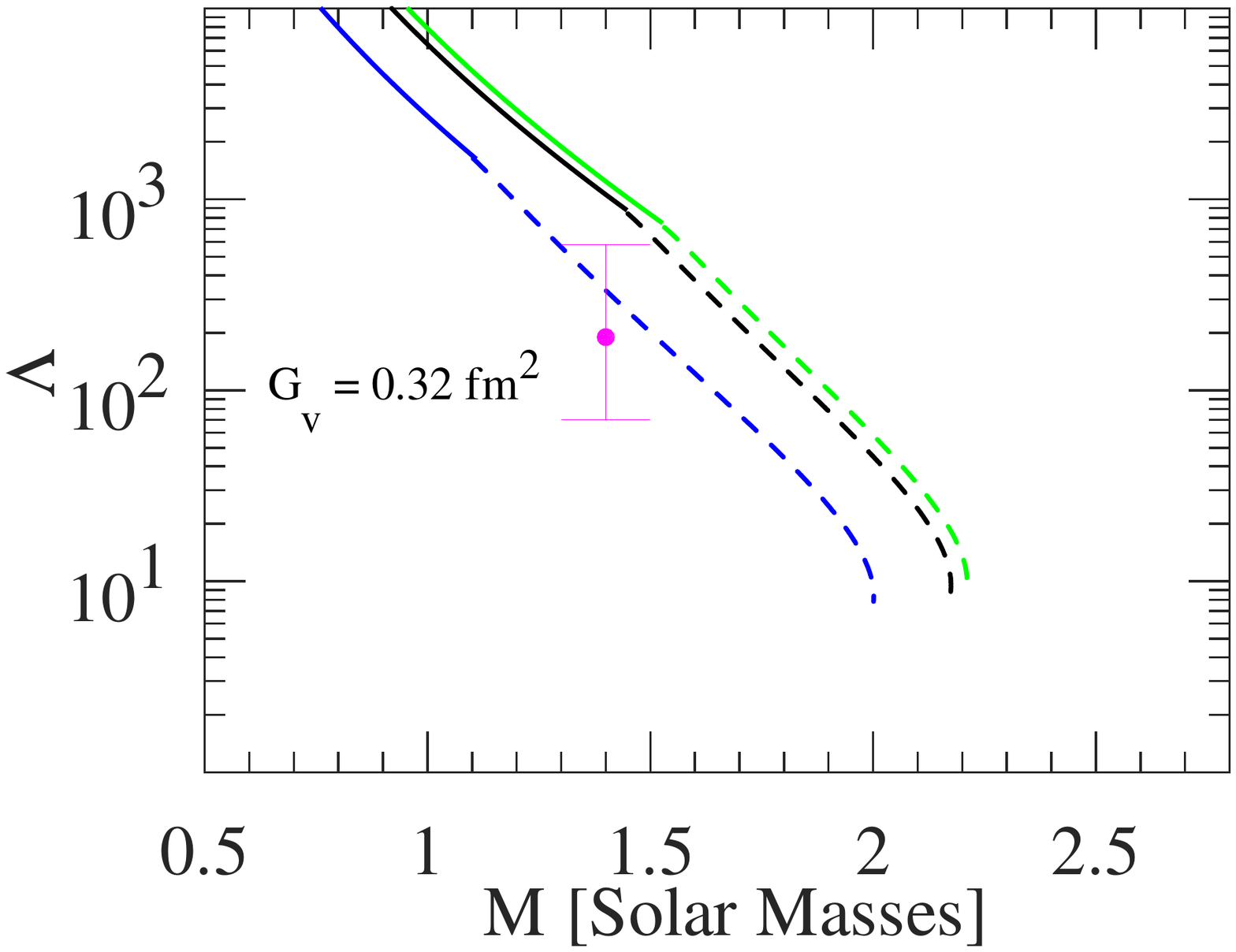} \\
\includegraphics[trim=0cm 6cm 0cm 6cm, clip, scale=0.33]{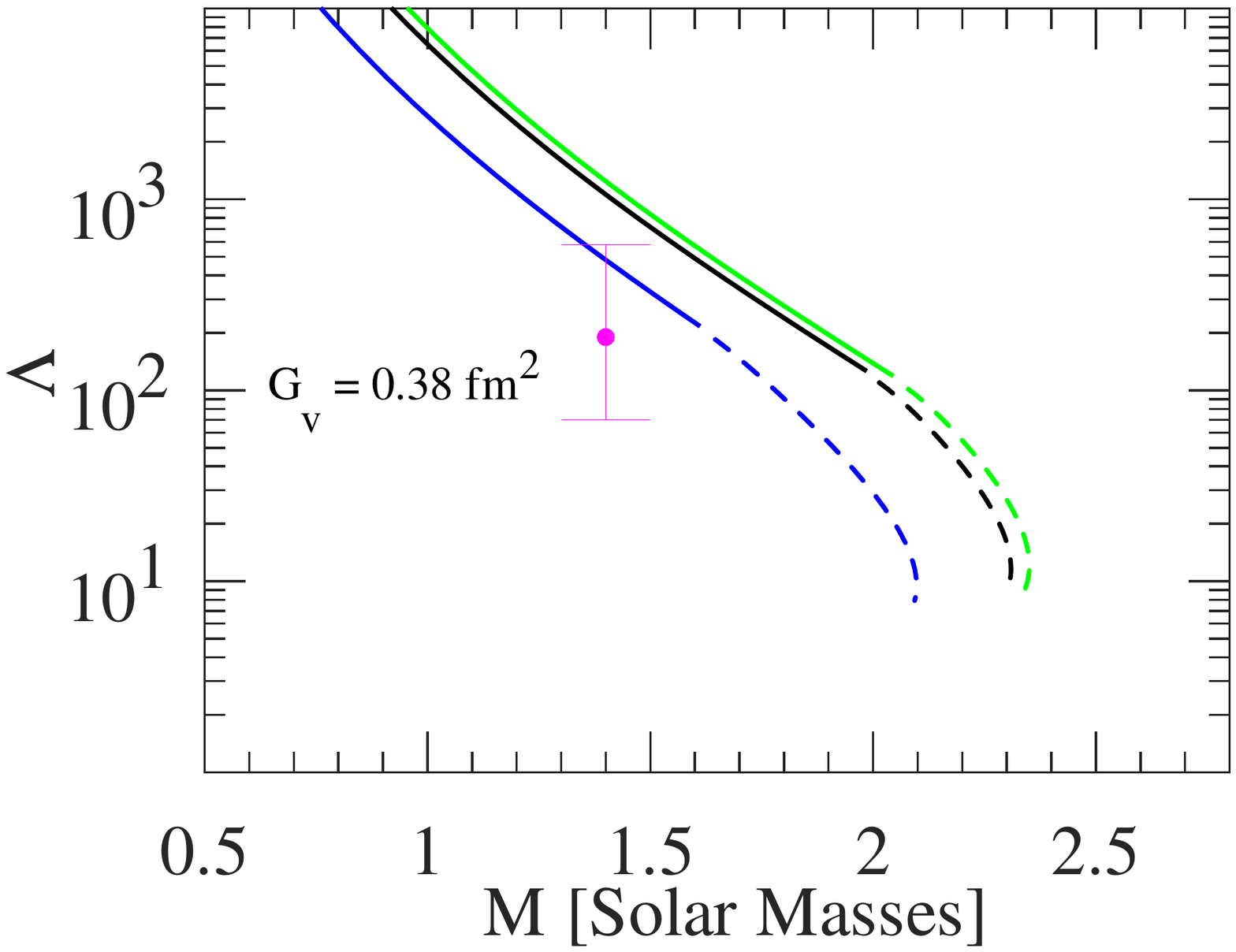} 
\includegraphics[trim=0cm 6cm 0cm 6cm, clip, scale=0.33]{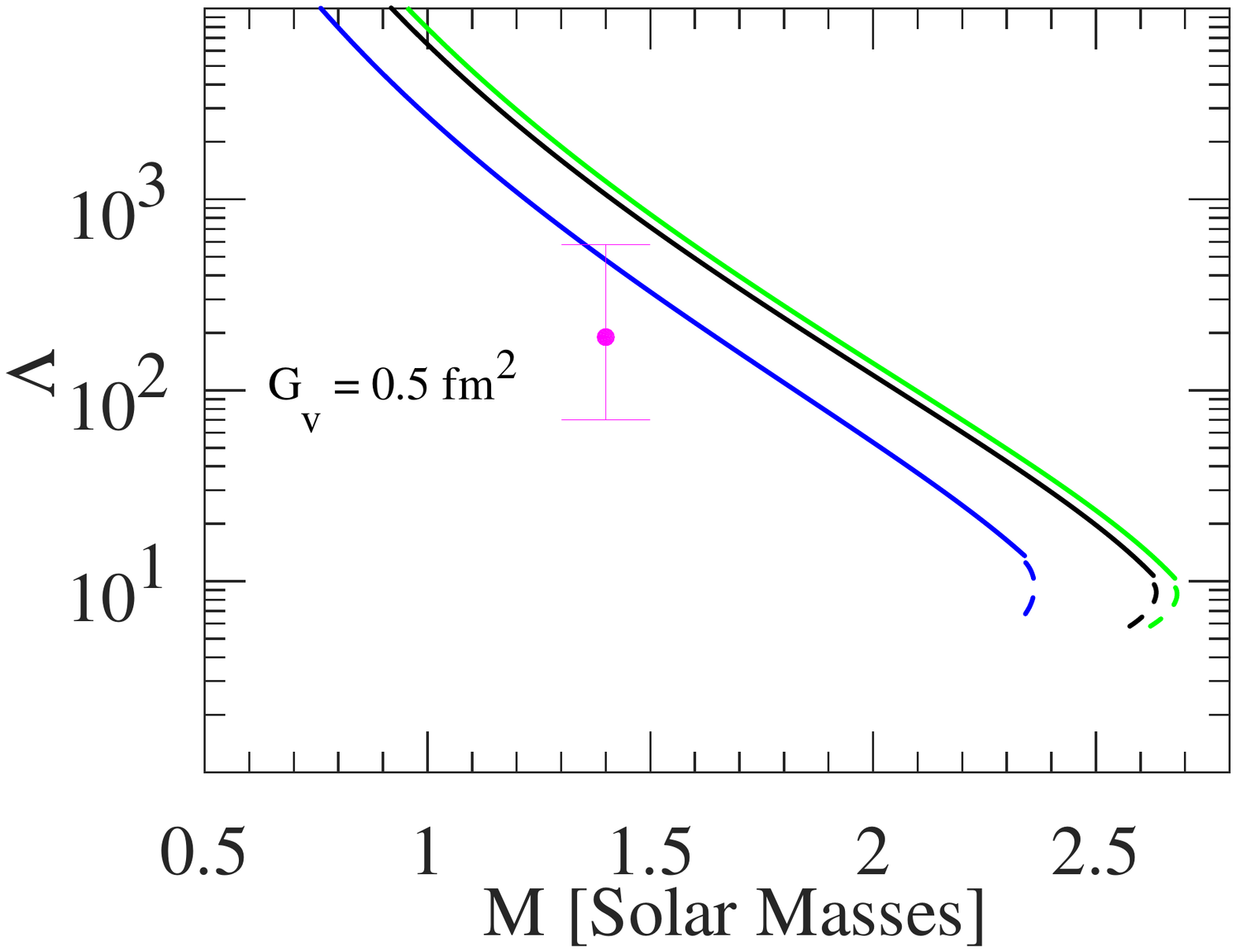}
\caption{Dimensionless tidal deformability $\Lambda$ as a function of the gravitational mass for the hybrid EoSs shown in Fig.~\ref{pt}. In the cases where $G_v = 0.2 \mbox{fm}^{2}$ and $G_v = 0.32 \mbox{fm}^{2}$ we can see that hybrid branch fulfils the constraint from the GW170817 event. On the other hand, the cases where $G_v = 0.38 \mbox{fm}^{2}$ and $G_v = 0.5 \mbox{fm}^{2}$ only hadronic branch, when $k_f = 0.04$, fulfills the GW170817 event.}
\label{MR2}
%\end{center}
\end{figure*}

In Fig.~\ref{MR2}, we show the dimensionless deformability parameter $\Lambda$ as a function of the mass. 
The first aspect that we can note is a decrease in $\Lambda$ with an increase in $k_F^{\mbox{\tiny DM}}$ for the same mass. This result was also verified for purely hadronic stars, see~\cite{dmnosso2}, for instance. On the other hand, it is evident that $\Lambda$ increases with the increase of $G_V$. The same results were observed in~\cite{ferrer}, in which authors studied strange stars formed by quark matter in the color-flavor-locked phase of color superconductivity, described by a Nambu-Jona-Lasinio type model with gluon contribution and vector interaction channel. Specially, notice that for the cases $G_V = 0.2~\mbox{fm}^{2}$ and $G_V = 0.32~\mbox{fm}^{2}$ the hybrid sector satisfies the GW170817 constraint, while in the other diagrams, we see that only hadronic stars present this feature.
%for some values of $k_F^{\mbox{\tiny DM}}$.

Furthermore, it is important to say that in compact star with just one phase, the larger the mass the smaller the tidal deformability for mechanical stable stars.~\cite{chatziioannou}. However, from Fig.~\ref{MR2} we see that $\Lambda$ can change its behaviour at larger masses because when we consider a slow phase-transition between hadron and quarks the last stable star  can be beyond the maximum mass. This behaviour is more evident in the case $G_V = 0.5 \mbox{ fm}^2$.

\begin{figure}
\centering
\includegraphics[scale=0.31]{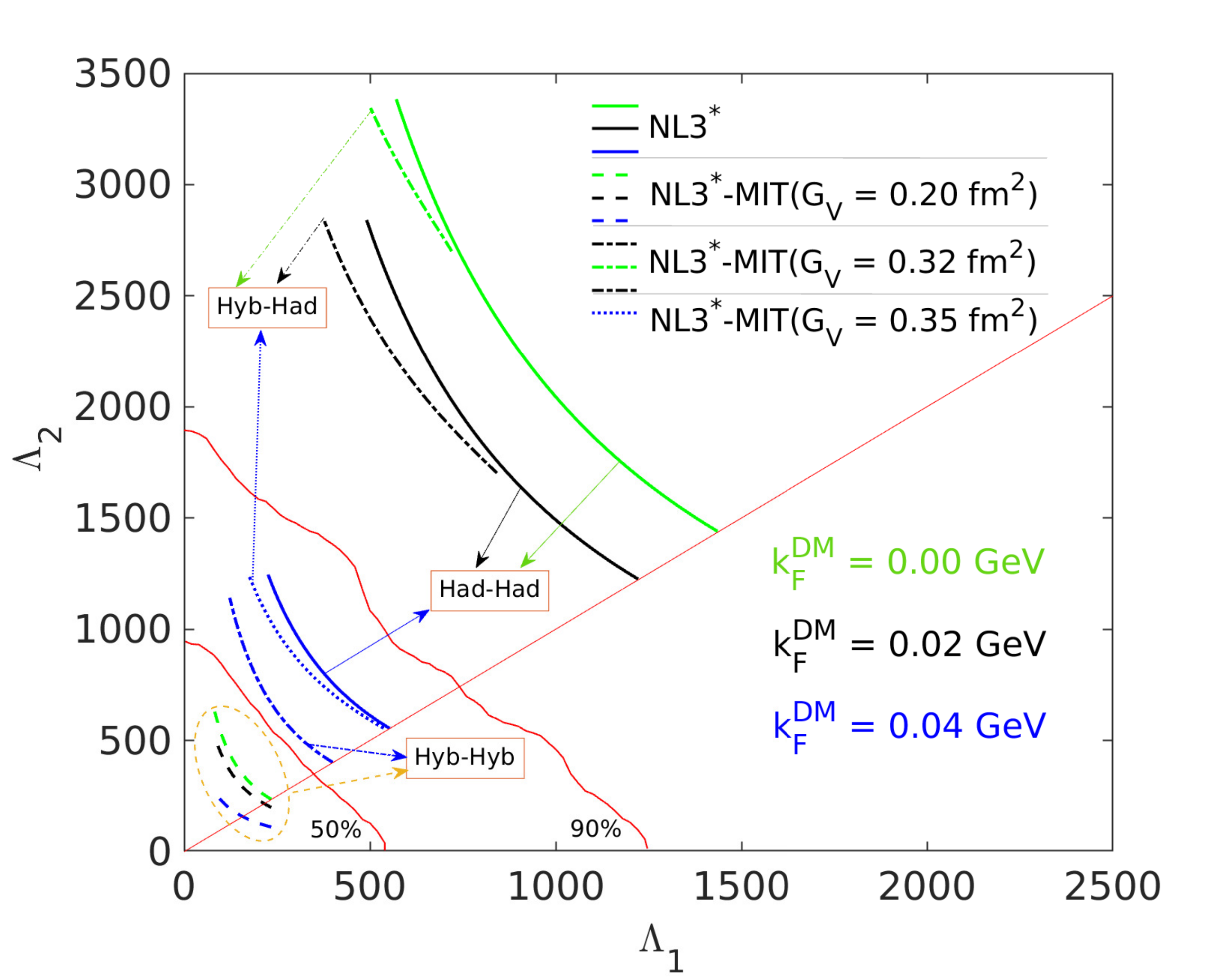}
\caption{Dimensionless tidal deformability parameters, $\Lambda_1$-$\Lambda_2$, for binary compact star mergers computed using the chirp mass of  the GW170817 event. The diagonal red line indicates the $\Lambda_{1}=\Lambda_{2}$ boundary and the other red lines denote the $50 \%$ and $90 \%$ confidence levels determined by the event GW170817. Different colors mean different values of $k_F^{\mbox{\tiny DM}}$, and for each kind of line we find a value of $G_V$ (dashed or dash-dotted lines), or pure hadronic modes (full line).}
\label{MR3}
\end{figure}

In Fig.~\ref{MR3}, we can see the relation between tidal deformability parameters, $\Lambda_1$-$\Lambda_2$, for binary compact star mergers computed using the chirp mass of  the GW170817 event:
\begin{equation}
M_C = (M_1 M_2)^{3/5} / (M_1+M_2)^{1 / 5} = 1.188 ~ {\rm M}_{\odot},   
\end{equation}
and the ratio $q = M_1/M_2$ in the range (0.7 - 1.0). In this way we can determine that the masses of the binary system vary in the ranges $1.36 ~{\rm M}_{\odot} < M_1  < 1.64~{\rm M}_{\odot}$ and  $1.14~{\rm M}_{\odot} < M_2  < 1.36~ {\rm M}_{\odot}$.

As it can be seen in Fig.~\ref{MR3} there are three different situations, namely: 
\begin{itemize}
    \item[i)] mergers composed of two purely hadronic stars (Had-Had);
    \item[ii)] mergers composed of pairs of hybrid stars (\mbox{Hyb-Hyb});   
    \item[iii)] mergers composed of a hybrid star and a hadronic star (Hyb-Had).
\end{itemize}

 Note that only one merger with two purely hadronic stars lies inside the $90\%$ confidence region of GW170817. Furthermore, it is possible to see that there are three %parametrizations
 cases where the binary systems are composed of a pair of  hybrid-hadronic stars. However, only the case with $G_V = 0.35 ~\mbox{fm}^{2}$ is  found inside the observable region of LIGO/Virgo. In the same diagram we find four mergers with two hybrid stars inside the $90\%$ confidence region, three of which lie inside the $50\%$ region. Finally, it is important to emphasize that the increase of $k_F^{\mbox{\tiny DM}}$ tends to move the curves inside the confidence regions of LIGO/Virgo. That is clear in all cases shown in Fig.~\ref{MR3}. Note also that all models analysed in $\Lambda_1-\Lambda_2$ diagram result in mass-radius sequences in agreement with constraints from the NICER observations.

Finally, in Table~\ref{TL3}, we can see a compilation of some of the most relevant results that we have obtained with this work. 
\begin{table*}
\setlength{\tabcolsep}{2pt}
\renewcommand{\arraystretch}{1.5}
\caption{Values of the maximum mass, $M_{max}$, radii of the maximum mass, $R$, the mass and radii of the last stable star, $M_{last}$ and $R_{last}$, the mass of the first hybrid star in the sequence, $M_{min}$, and the values of the radii and tidal deformability, $R_{1.4}$ and $\Lambda_{1.4}$, for different coupling constants of the vector MTI bag model, $G_V$, and dark matter Fermi momentum, $k_F^{\mbox{\tiny DM}}$.}
\begin{center}
\begin{tabular}{|c|c|c|c|c|c|c|c|c|}
\hline 
  \ \ $G_V$(fm$^{2}$) \ \  & \ \ $k_F^{\mbox{\tiny DM}}$ (GeV) \ \ & \ \ $M_{max}$ ($M_\odot$) \ \ & \ \ $R$(km) \ \   & \ \  $M_{last}$ ($M_\odot$) \ \ & \ \ R$_{last}$(km) \ \ & \ \ $M_{min}$($M_\odot$) \ \ & 
 \ \ $R_{1.4}$ (km) \ \  & \ \ $\Lambda_{1.4}$\\
  \hline \hline
0.2  & 0.00 & 2.15 & 11.47 & 2.15  & 10.25 & 0.64 & 12.91& 201 \\
 \hline
0.2  & 0.02 & 2.12 & 11.17 & 2.12  & 11.17 & 0.57 &12.43& 170 \\
 \hline
0.2  & 0.04 & 1.97 & 10.05 & 1.97  & 10.05 & 0.33 &10.72& 96  \\
 \hline \hline
0.32 & 0.00 & 2.21 & 12.35 & 2.20  & 12.14 & 1.53 &15.44&1250 \\
 \hline
0.32 & 0.02 & 2.18 & 12.00 & 2.17  & 11.83 & 1.44 & 14.76&1059 \\
 \hline
0.32 & 0.04 & 2.00 & 10.45 & 2.00  & 9.58 & 1.10 &11.74&332 \\ 
\hline \hline
0.35 & 0.00 & 2.28 & 12.73 & 2.27  & 12.40 & 1.81 &15.44& 1250 \\
 \hline
0.35 & 0.02 & 2.24 & 12.36 & 2.23  & 12.09 & 1.74 &14.76&1059 \\
 \hline
0.35 & 0.04 & 2.05 & 10.45 & 2.04  & 9.58 & 1.38 &12.08&458 \\ 
\hline \hline
0.38 & 0.00 & 2.35 & 13.04 & 2.34  & 12.65 & 2.07 &15.44& 1250 \\
 \hline
0.38 & 0.02 & 2.31 & 12.76 & 2.30  & 12.33 & 2.01 &14.76&1059 \\
 \hline
0.38 & 0.04 & 2.10 & 11.02 & 2.09  & 10.81 & 1.64 &12.17&483 \\ 
\hline \hline
0.5 & 0.00 & 2.68 & 13.99 & 2.62   & 13.20 & 2.68 &15.44& 1250 \\
 \hline
0.5 & 0.02 & 2.64 & 13.58 & 2.57   & 12.90 & 2.63 &14.76&1059 \\
 \hline
0.5 & 0.04 & 2.36 & 11.97 & 2.32   & 11.33 & 2.34 &12.17&483 \\ 
\hline
\end{tabular}
\label{TL3}
\end{center}
\end{table*}
As it is clear in Table~\ref{TL3} the radii and tidal deformability of the canonical stars, $R_{1.4}$ and $\Lambda_{1.4}$, decrease with $k_F^{DM}$ and, with few exceptions - discussed next, increase with $G_V$. In some cases, the stars satisfy the GW170817 constraint. Note also that for some parametrizations the values of $R_{1.4}$ and $\Lambda_{1.4}$ are repeated. That occurs because in most cases the star with $1.4M_\odot$ appears in the hadronic sector. It is important to mention that the last stable star mass values, $M_{last}$, computed considering slow phase transitions, remained close to the maximum mass, $M_{max}$, excepted in the case of $G_V = 0.5 ~{\rm fm}^{2}$, where we can see the greatest differences between those values. This fact is directly related to the value of the gap, $\Delta \epsilon$, in Table~\ref{table2}. The greater the gap, the greater the sequence of the stable stars beyond the maximum mass~\cite{rad2}.  

\section{Summary and concluding remarks}

In this paper, we have studied hybrid stars by considering hadrons and quarks admixed with dark matter. In the hadronic side we have used a particular parametrization (NL3*) consistent with some finite nuclei quantities, and for the effective quark model we have considered a version of the MIT bag model in which a vector channel type interaction is taken into account. As a consequence, we have investigated how star properties vary by changing both, the strength of this vector interaction as well as the Fermi momentum of the dark particle assumed as the DM candidate, namely, the lightest neutralino. Regarding this investigation, our main conclusions are the following,
\begin{itemize}

    \item  Dark matter does not influence the phase transition point, i.e, the ($\mu_t$, $P_t$) pair. Furthermore, the already known feature  \cite{Lopes_2022} is again obtained: the strength of the vector interaction influences the transition from hadronic to quark matter such that the larger the value of $G_V$, the higher the pressure and chemical potential at the transition point. However, as it always shifts the energy density towards higher values and  does not change the pressure, the energy density gap is shifted with the increase of the dark matter Fermi momentum.
    
    \item Dark matter makes the maximum stellar mass smaller and deviates the radii of the whole star family to smaller values, in agreement with previous studies~\cite{rmfdm2,rmfdm3,abdul,rmfdm6,rmfdm8,rmfdm10,rmfdm11}, but values compatible with recent NICER observational data are easily obtained. In summary, we show that this particular effect is also verified for hybrid stars, constructed with admixture of DM on both sides. 
    
    \item The observational data predicted by LIGO/Virgo Collaboration concerning the GW170817 event are more easily attained with the inclusion of dark matter. This feature was also observed in the construction of purely hadronic stars with DM included, as the reader can verify in~\cite{dmnosso2,rmfdm3,feebly}. Furthermore, we found hybrid-hybrid, hybrid-hadronic and hadronic-hadronic configurations for the stars of the binary system, both containing DM, consistent with the $\Lambda_1\times\Lambda_2$ region.

    \item Dark matter favours the appearance of large quark cores { and reduces the mass of the first hybrid star ($M_{min}$)}.

\end{itemize}

\section*{ACKNOWLEDGMENTS}
This work is a part of the project INCT-FNA proc. No. 464898/2014-5. It is also supported by Conselho Nacional de Desenvolvimento Cient\'ifico e Tecnol\'ogico (CNPq) under Grants No. 312410/2020-4 (O.L.), No. 308528/2021-2 (M.D.) and No. 303490/2021-7 (D.P.M.). O.L., M.D. and C.H.L. also acknowledge Funda\c{c}\~ao de Amparo \`a Pesquisa do Estado de S\~ao Paulo (FAPESP) under Thematic Project 2017/05660-0 and Grant No. 2020/05238-9. O.~L. is also supported by FAPESP under Grant No. 2022/03575-3 (BPE). The authors gratefully acknowledge A.~S.~Schneider for stimulating discussions and valuable comments.

\bibliography{references}

\begin{thebibliography}{56}%
\makeatletter
\providecommand \@ifxundefined [1]{%
 \@ifx{#1\undefined}
}%
\providecommand \@ifnum [1]{%
 \ifnum #1\expandafter \@firstoftwo
 \else \expandafter \@secondoftwo
 \fi
}%
\providecommand \@ifx [1]{%
 \ifx #1\expandafter \@firstoftwo
 \else \expandafter \@secondoftwo
 \fi
}%
\providecommand \natexlab [1]{#1}%
\providecommand \enquote  [1]{``#1''}%
\providecommand \bibnamefont  [1]{#1}%
\providecommand \bibfnamefont [1]{#1}%
\providecommand \citenamefont [1]{#1}%
\providecommand \href@noop [0]{\@secondoftwo}%
\providecommand \href [0]{\begingroup \@sanitize@url \@href}%
\providecommand \@href[1]{\@@startlink{#1}\@@href}%
\providecommand \@@href[1]{\endgroup#1\@@endlink}%
\providecommand \@sanitize@url [0]{\catcode `\\12\catcode `\$12\catcode
  `\&12\catcode `\#12\catcode `\^12\catcode `\_12\catcode `\%12\relax}%
\providecommand \@@startlink[1]{}%
\providecommand \@@endlink[0]{}%
\providecommand \url  [0]{\begingroup\@sanitize@url \@url }%
\providecommand \@url [1]{\endgroup\@href {#1}{\urlprefix }}%
\providecommand \urlprefix  [0]{URL }%
\providecommand \Eprint [0]{\href }%
\providecommand \doibase [0]{https://doi.org/}%
\providecommand \selectlanguage [0]{\@gobble}%
\providecommand \bibinfo  [0]{\@secondoftwo}%
\providecommand \bibfield  [0]{\@secondoftwo}%
\providecommand \translation [1]{[#1]}%
\providecommand \BibitemOpen [0]{}%
\providecommand \bibitemStop [0]{}%
\providecommand \bibitemNoStop [0]{.\EOS\space}%
\providecommand \EOS [0]{\spacefactor3000\relax}%
\providecommand \BibitemShut  [1]{\csname bibitem#1\endcsname}%
\let\auto@bib@innerbib\@empty
%</preamble>
\bibitem [{\citenamefont {Bertone}\ and\ \citenamefont
  {Hooper}(2018)}]{bertone}%
  \BibitemOpen
  \bibfield  {author} {\bibinfo {author} {\bibfnamefont {G.}~\bibnamefont
  {Bertone}}\ and\ \bibinfo {author} {\bibfnamefont {D.}~\bibnamefont
  {Hooper}},\ }\href {https://doi.org/10.1103/RevModPhys.90.045002} {\bibfield
  {journal} {\bibinfo  {journal} {Rev. Mod. Phys.}\ }\textbf {\bibinfo {volume}
  {90}},\ \bibinfo {pages} {045002} (\bibinfo {year} {2018})}\BibitemShut
  {NoStop}%
\bibitem [{\citenamefont {Workman}\ and\ \citenamefont {Others}(2022)}]{pdg22}%
  \BibitemOpen
  \bibfield  {author} {\bibinfo {author} {\bibfnamefont {R.~L.}\ \bibnamefont
  {Workman}}\ and\ \bibinfo {author} {\bibnamefont {Others}} (\bibinfo
  {collaboration} {Particle Data Group}),\ }\href
  {https://doi.org/10.1093/ptep/ptac097} {\bibfield  {journal} {\bibinfo
  {journal} {PTEP}\ }\textbf {\bibinfo {volume} {2022}},\ \bibinfo {pages}
  {083C01} (\bibinfo {year} {2022})}\BibitemShut {NoStop}%
\bibitem [{\citenamefont {Das}\ \emph {et~al.}(2022{\natexlab{a}})\citenamefont
  {Das}, \citenamefont {Kumar}, \citenamefont {Kumar},\ and\ \citenamefont
  {Patra}}]{rmfdm13}%
  \BibitemOpen
  \bibfield  {author} {\bibinfo {author} {\bibfnamefont {H.~C.}\ \bibnamefont
  {Das}}, \bibinfo {author} {\bibfnamefont {A.}~\bibnamefont {Kumar}}, \bibinfo
  {author} {\bibfnamefont {B.}~\bibnamefont {Kumar}},\ and\ \bibinfo {author}
  {\bibfnamefont {S.~K.}\ \bibnamefont {Patra}},\ }\bibfield  {journal}
  {\bibinfo  {journal} {Galaxies}\ }\textbf {\bibinfo {volume} {10}},\ \href
  {https://doi.org/10.3390/galaxies10010014} {10.3390/galaxies10010014}
  (\bibinfo {year} {2022}{\natexlab{a}})\BibitemShut {NoStop}%
\bibitem [{\citenamefont {Das}\ \emph {et~al.}(2022{\natexlab{b}})\citenamefont
  {Das}, \citenamefont {Malik},\ and\ \citenamefont {Nayak}}]{rmfdm1}%
  \BibitemOpen
  \bibfield  {author} {\bibinfo {author} {\bibfnamefont {A.}~\bibnamefont
  {Das}}, \bibinfo {author} {\bibfnamefont {T.}~\bibnamefont {Malik}},\ and\
  \bibinfo {author} {\bibfnamefont {A.~C.}\ \bibnamefont {Nayak}},\ }\href
  {https://doi.org/10.1103/PhysRevD.105.123034} {\bibfield  {journal} {\bibinfo
   {journal} {Phys. Rev. D}\ }\textbf {\bibinfo {volume} {105}},\ \bibinfo
  {pages} {123034} (\bibinfo {year} {2022}{\natexlab{b}})}\BibitemShut
  {NoStop}%
\bibitem [{\citenamefont {Panotopoulos}\ and\ \citenamefont
  {Lopes}(2017{\natexlab{a}})}]{rmfdm2}%
  \BibitemOpen
  \bibfield  {author} {\bibinfo {author} {\bibfnamefont {G.}~\bibnamefont
  {Panotopoulos}}\ and\ \bibinfo {author} {\bibfnamefont {I.}~\bibnamefont
  {Lopes}},\ }\href {https://doi.org/10.1103/PhysRevD.96.083004} {\bibfield
  {journal} {\bibinfo  {journal} {Phys. Rev. D}\ }\textbf {\bibinfo {volume}
  {96}},\ \bibinfo {pages} {083004} (\bibinfo {year}
  {2017}{\natexlab{a}})}\BibitemShut {NoStop}%
\bibitem [{\citenamefont {Das}\ \emph {et~al.}(2019)\citenamefont {Das},
  \citenamefont {Malik},\ and\ \citenamefont {Nayak}}]{rmfdm3}%
  \BibitemOpen
  \bibfield  {author} {\bibinfo {author} {\bibfnamefont {A.}~\bibnamefont
  {Das}}, \bibinfo {author} {\bibfnamefont {T.}~\bibnamefont {Malik}},\ and\
  \bibinfo {author} {\bibfnamefont {A.~C.}\ \bibnamefont {Nayak}},\ }\href
  {https://doi.org/10.1103/PhysRevD.99.043016} {\bibfield  {journal} {\bibinfo
  {journal} {Phys. Rev. D}\ }\textbf {\bibinfo {volume} {99}},\ \bibinfo
  {pages} {043016} (\bibinfo {year} {2019})}\BibitemShut {NoStop}%
\bibitem [{\citenamefont {Quddus}\ \emph {et~al.}(2020)\citenamefont {Quddus},
  \citenamefont {Panotopoulos}, \citenamefont {Kumar}, \citenamefont {Ahmad},\
  and\ \citenamefont {Patra}}]{abdul}%
  \BibitemOpen
  \bibfield  {author} {\bibinfo {author} {\bibfnamefont {A.}~\bibnamefont
  {Quddus}}, \bibinfo {author} {\bibfnamefont {G.}~\bibnamefont
  {Panotopoulos}}, \bibinfo {author} {\bibfnamefont {B.}~\bibnamefont {Kumar}},
  \bibinfo {author} {\bibfnamefont {S.}~\bibnamefont {Ahmad}},\ and\ \bibinfo
  {author} {\bibfnamefont {S.~K.}\ \bibnamefont {Patra}},\ }\href
  {https://doi.org/10.1088/1361-6471/ab9d36} {\bibfield  {journal} {\bibinfo
  {journal} {Journal of Physics G: Nuclear and Particle Physics}\ }\textbf
  {\bibinfo {volume} {47}},\ \bibinfo {pages} {095202} (\bibinfo {year}
  {2020})}\BibitemShut {NoStop}%
\bibitem [{\citenamefont {Das}\ \emph {et~al.}(2020)\citenamefont {Das},
  \citenamefont {Kumar}, \citenamefont {Kumar}, \citenamefont {Biswal},
  \citenamefont {Nakatsukasa}, \citenamefont {Li},\ and\ \citenamefont
  {Patra}}]{rmfdm6}%
  \BibitemOpen
  \bibfield  {author} {\bibinfo {author} {\bibfnamefont {H.~C.}\ \bibnamefont
  {Das}}, \bibinfo {author} {\bibfnamefont {A.}~\bibnamefont {Kumar}}, \bibinfo
  {author} {\bibfnamefont {B.}~\bibnamefont {Kumar}}, \bibinfo {author}
  {\bibfnamefont {S.~K.}\ \bibnamefont {Biswal}}, \bibinfo {author}
  {\bibfnamefont {T.}~\bibnamefont {Nakatsukasa}}, \bibinfo {author}
  {\bibfnamefont {A.}~\bibnamefont {Li}},\ and\ \bibinfo {author}
  {\bibfnamefont {S.~K.}\ \bibnamefont {Patra}},\ }\href
  {https://doi.org/10.1093/mnras/staa1435} {\bibfield  {journal} {\bibinfo
  {journal} {Monthly Notices of the Royal Astronomical Society}\ }\textbf
  {\bibinfo {volume} {495}},\ \bibinfo {pages} {4893} (\bibinfo {year}
  {2020})}\BibitemShut {NoStop}%
\bibitem [{\citenamefont {Das}\ \emph {et~al.}(2021{\natexlab{a}})\citenamefont
  {Das}, \citenamefont {Kumar},\ and\ \citenamefont {Patra}}]{rmfdm11}%
  \BibitemOpen
  \bibfield  {author} {\bibinfo {author} {\bibfnamefont {H.~C.}\ \bibnamefont
  {Das}}, \bibinfo {author} {\bibfnamefont {A.}~\bibnamefont {Kumar}},\ and\
  \bibinfo {author} {\bibfnamefont {S.~K.}\ \bibnamefont {Patra}},\ }\href
  {https://doi.org/10.1103/PhysRevD.104.063028} {\bibfield  {journal} {\bibinfo
   {journal} {Phys. Rev. D}\ }\textbf {\bibinfo {volume} {104}},\ \bibinfo
  {pages} {063028} (\bibinfo {year} {2021}{\natexlab{a}})}\BibitemShut
  {NoStop}%
\bibitem [{\citenamefont {Das}\ \emph {et~al.}(2021{\natexlab{b}})\citenamefont
  {Das}, \citenamefont {Kumar}, \citenamefont {Biswal},\ and\ \citenamefont
  {Patra}}]{rmfdm10}%
  \BibitemOpen
  \bibfield  {author} {\bibinfo {author} {\bibfnamefont {H.~C.}\ \bibnamefont
  {Das}}, \bibinfo {author} {\bibfnamefont {A.}~\bibnamefont {Kumar}}, \bibinfo
  {author} {\bibfnamefont {S.~K.}\ \bibnamefont {Biswal}},\ and\ \bibinfo
  {author} {\bibfnamefont {S.~K.}\ \bibnamefont {Patra}},\ }\href
  {https://doi.org/10.1103/PhysRevD.104.123006} {\bibfield  {journal} {\bibinfo
   {journal} {Phys. Rev. D}\ }\textbf {\bibinfo {volume} {104}},\ \bibinfo
  {pages} {123006} (\bibinfo {year} {2021}{\natexlab{b}})}\BibitemShut
  {NoStop}%
\bibitem [{\citenamefont {Das}\ \emph {et~al.}(2021{\natexlab{c}})\citenamefont
  {Das}, \citenamefont {Kumar},\ and\ \citenamefont {Patra}}]{rmfdm8}%
  \BibitemOpen
  \bibfield  {author} {\bibinfo {author} {\bibfnamefont {H.~C.}\ \bibnamefont
  {Das}}, \bibinfo {author} {\bibfnamefont {A.}~\bibnamefont {Kumar}},\ and\
  \bibinfo {author} {\bibfnamefont {S.~K.}\ \bibnamefont {Patra}},\ }\href
  {https://doi.org/10.1093/mnras/stab2387} {\bibfield  {journal} {\bibinfo
  {journal} {Monthly Notices of the Royal Astronomical Society}\ }\textbf
  {\bibinfo {volume} {507}},\ \bibinfo {pages} {4053} (\bibinfo {year}
  {2021}{\natexlab{c}})}\BibitemShut {NoStop}%
\bibitem [{\citenamefont {Das}\ \emph {et~al.}(2021{\natexlab{d}})\citenamefont
  {Das}, \citenamefont {Kumar}, \citenamefont {Kumar}, \citenamefont {Biswal},\
  and\ \citenamefont {Patra}}]{rmfdm7}%
  \BibitemOpen
  \bibfield  {author} {\bibinfo {author} {\bibfnamefont {H.}~\bibnamefont
  {Das}}, \bibinfo {author} {\bibfnamefont {A.}~\bibnamefont {Kumar}}, \bibinfo
  {author} {\bibfnamefont {B.}~\bibnamefont {Kumar}}, \bibinfo {author}
  {\bibfnamefont {S.}~\bibnamefont {Biswal}},\ and\ \bibinfo {author}
  {\bibfnamefont {S.}~\bibnamefont {Patra}},\ }\href
  {https://doi.org/10.1088/1475-7516/2021/01/007} {\bibfield  {journal}
  {\bibinfo  {journal} {Journal of Cosmology and Astroparticle Physics}\
  }\textbf {\bibinfo {volume} {2021}}\bibinfo  {number} { (01)},\ \bibinfo
  {pages} {007}}\BibitemShut {NoStop}%
\bibitem [{\citenamefont {Kumar}\ \emph {et~al.}(2022)\citenamefont {Kumar},
  \citenamefont {Das},\ and\ \citenamefont {Patra}}]{rmfdm12}%
  \BibitemOpen
\bibfield  {number} {  }\bibfield  {author} {\bibinfo {author} {\bibfnamefont
  {A.}~\bibnamefont {Kumar}}, \bibinfo {author} {\bibfnamefont {H.~C.}\
  \bibnamefont {Das}},\ and\ \bibinfo {author} {\bibfnamefont {S.~K.}\
  \bibnamefont {Patra}},\ }\href {https://doi.org/10.1093/mnras/stac1013}
  {\bibfield  {journal} {\bibinfo  {journal} {Monthly Notices of the Royal
  Astronomical Society}\ }\textbf {\bibinfo {volume} {513}},\ \bibinfo {pages}
  {1820} (\bibinfo {year} {2022})}\BibitemShut {NoStop}%
\bibitem [{\citenamefont {Louren\ifmmode~\mbox{\c{c}}\else \c{c}\fi{}o}\ \emph
  {et~al.}(2022)\citenamefont {Louren\ifmmode~\mbox{\c{c}}\else \c{c}\fi{}o},
  \citenamefont {Frederico},\ and\ \citenamefont {Dutra}}]{dmnosso1}%
  \BibitemOpen
  \bibfield  {author} {\bibinfo {author} {\bibfnamefont {O.}~\bibnamefont
  {Louren\ifmmode~\mbox{\c{c}}\else \c{c}\fi{}o}}, \bibinfo {author}
  {\bibfnamefont {T.}~\bibnamefont {Frederico}},\ and\ \bibinfo {author}
  {\bibfnamefont {M.}~\bibnamefont {Dutra}},\ }\href
  {https://doi.org/10.1103/PhysRevD.105.023008} {\bibfield  {journal} {\bibinfo
   {journal} {Phys. Rev. D}\ }\textbf {\bibinfo {volume} {105}},\ \bibinfo
  {pages} {023008} (\bibinfo {year} {2022})}\BibitemShut {NoStop}%
\bibitem [{\citenamefont {Lourenço}\ \emph {et~al.}(2022)\citenamefont
  {Lourenço}, \citenamefont {Lenzi}, \citenamefont {Frederico},\ and\
  \citenamefont {Dutra}}]{dmnosso2}%
  \BibitemOpen
  \bibfield  {author} {\bibinfo {author} {\bibfnamefont {O.}~\bibnamefont
  {Lourenço}}, \bibinfo {author} {\bibfnamefont {C.~H.}\ \bibnamefont
  {Lenzi}}, \bibinfo {author} {\bibfnamefont {T.}~\bibnamefont {Frederico}},\
  and\ \bibinfo {author} {\bibfnamefont {M.}~\bibnamefont {Dutra}},\ }\href
  {https://doi.org/10.1103/PhysRevD.106.043010} {\bibfield  {journal} {\bibinfo
   {journal} {Phys. Rev. D}\ }\textbf {\bibinfo {volume} {106}},\ \bibinfo
  {pages} {043010} (\bibinfo {year} {2022})}\BibitemShut {NoStop}%
\bibitem [{\citenamefont {Dutra}\ \emph {et~al.}(2022)\citenamefont {Dutra},
  \citenamefont {Lenzi},\ and\ \citenamefont {Lourenço}}]{dmnosso3}%
  \BibitemOpen
  \bibfield  {author} {\bibinfo {author} {\bibfnamefont {M.}~\bibnamefont
  {Dutra}}, \bibinfo {author} {\bibfnamefont {C.~H.}\ \bibnamefont {Lenzi}},\
  and\ \bibinfo {author} {\bibfnamefont {O.}~\bibnamefont {Lourenço}},\ }\href
  {https://doi.org/10.1093/mnras/stac2986} {\bibfield  {journal} {\bibinfo
  {journal} {Monthly Notices of the Royal Astronomical Society}\ }\textbf
  {\bibinfo {volume} {517}},\ \bibinfo {pages} {4265} (\bibinfo {year}
  {2022})}\BibitemShut {NoStop}%
\bibitem [{\citenamefont {Dengler}\ \emph {et~al.}(2022)\citenamefont
  {Dengler}, \citenamefont {Schaffner-Bielich},\ and\ \citenamefont
  {Tolos}}]{laura-tolos}%
  \BibitemOpen
  \bibfield  {author} {\bibinfo {author} {\bibfnamefont {Y.}~\bibnamefont
  {Dengler}}, \bibinfo {author} {\bibfnamefont {J.}~\bibnamefont
  {Schaffner-Bielich}},\ and\ \bibinfo {author} {\bibfnamefont
  {L.}~\bibnamefont {Tolos}},\ }\href
  {https://doi.org/10.1103/PhysRevD.105.043013} {\bibfield  {journal} {\bibinfo
   {journal} {Phys. Rev. D}\ }\textbf {\bibinfo {volume} {105}},\ \bibinfo
  {pages} {043013} (\bibinfo {year} {2022})}\BibitemShut {NoStop}%
\bibitem [{\citenamefont {Giangrandi}\ \emph {et~al.}(2022)\citenamefont
  {Giangrandi}, \citenamefont {Sagun}, \citenamefont {Ivanytskyi},
  \citenamefont {Providência},\ and\ \citenamefont {Dietrich}}]{sagun}%
  \BibitemOpen
  \bibfield  {author} {\bibinfo {author} {\bibfnamefont {E.}~\bibnamefont
  {Giangrandi}}, \bibinfo {author} {\bibfnamefont {V.}~\bibnamefont {Sagun}},
  \bibinfo {author} {\bibfnamefont {O.}~\bibnamefont {Ivanytskyi}}, \bibinfo
  {author} {\bibfnamefont {C.}~\bibnamefont {Providência}},\ and\ \bibinfo
  {author} {\bibfnamefont {T.}~\bibnamefont {Dietrich}},\ }\href
  {https://doi.org/10.48550/ARXIV.2209.10905} {\bibinfo {title} {The effects of
  self-interacting bosonic dark matter on neutron star properties}} (\bibinfo
  {year} {2022})\BibitemShut {NoStop}%
\bibitem [{\citenamefont {Lopes}\ and\ \citenamefont
  {Menezes}(2018)}]{Lopes_2018}%
  \BibitemOpen
  \bibfield  {author} {\bibinfo {author} {\bibfnamefont {L.}~\bibnamefont
  {Lopes}}\ and\ \bibinfo {author} {\bibfnamefont {D.}~\bibnamefont
  {Menezes}},\ }\href {https://doi.org/10.1088/1475-7516/2018/05/038}
  {\bibfield  {journal} {\bibinfo  {journal} {Journal of Cosmology and
  Astroparticle Physics}\ }\textbf {\bibinfo {volume} {2018}}\bibinfo  {number}
  { (05)},\ \bibinfo {pages} {038}}\BibitemShut {NoStop}%
\bibitem [{\citenamefont {Rafiei~Karkevandi}\ \emph {et~al.}(2022)\citenamefont
  {Rafiei~Karkevandi}, \citenamefont {Shakeri}, \citenamefont {Sagun},\ and\
  \citenamefont {Ivanytskyi}}]{karkevandi1}%
  \BibitemOpen
\bibfield  {number} {  }\bibfield  {author} {\bibinfo {author} {\bibfnamefont
  {D.}~\bibnamefont {Rafiei~Karkevandi}}, \bibinfo {author} {\bibfnamefont
  {S.}~\bibnamefont {Shakeri}}, \bibinfo {author} {\bibfnamefont
  {V.}~\bibnamefont {Sagun}},\ and\ \bibinfo {author} {\bibfnamefont
  {O.}~\bibnamefont {Ivanytskyi}},\ }\href
  {https://doi.org/10.1103/PhysRevD.105.023001} {\bibfield  {journal} {\bibinfo
   {journal} {Phys. Rev. D}\ }\textbf {\bibinfo {volume} {105}},\ \bibinfo
  {pages} {023001} (\bibinfo {year} {2022})}\BibitemShut {NoStop}%
\bibitem [{\citenamefont {Shakeri}\ and\ \citenamefont
  {Karkevandi}(2022)}]{karkevandi2}%
  \BibitemOpen
  \bibfield  {author} {\bibinfo {author} {\bibfnamefont {S.}~\bibnamefont
  {Shakeri}}\ and\ \bibinfo {author} {\bibfnamefont {D.~R.}\ \bibnamefont
  {Karkevandi}},\ }\href {https://doi.org/10.48550/ARXIV.2210.17308} {\bibinfo
  {title} {Bosonic dark matter in light of the nicer precise mass-radius
  measurements}} (\bibinfo {year} {2022})\BibitemShut {NoStop}%
\bibitem [{\citenamefont {Jiménez}\ and\ \citenamefont {Fraga}(2022)}]{fraga}%
  \BibitemOpen
  \bibfield  {author} {\bibinfo {author} {\bibfnamefont {J.~C.}\ \bibnamefont
  {Jiménez}}\ and\ \bibinfo {author} {\bibfnamefont {E.~S.}\ \bibnamefont
  {Fraga}},\ }\bibfield  {journal} {\bibinfo  {journal} {Universe}\ }\textbf
  {\bibinfo {volume} {8}},\ \href {https://doi.org/10.3390/universe8010034}
  {10.3390/universe8010034} (\bibinfo {year} {2022})\BibitemShut {NoStop}%
\bibitem [{\citenamefont {{Annala}}\ \emph {et~al.}(2020)\citenamefont
  {{Annala}}, \citenamefont {{Gorda}}, \citenamefont {{Kurkela}}, \citenamefont
  {{N{\"a}ttil{\"a}}},\ and\ \citenamefont {{Vuorinen}}}]{nature_2020}%
  \BibitemOpen
  \bibfield  {author} {\bibinfo {author} {\bibfnamefont {E.}~\bibnamefont
  {{Annala}}}, \bibinfo {author} {\bibfnamefont {T.}~\bibnamefont {{Gorda}}},
  \bibinfo {author} {\bibfnamefont {A.}~\bibnamefont {{Kurkela}}}, \bibinfo
  {author} {\bibfnamefont {J.}~\bibnamefont {{N{\"a}ttil{\"a}}}},\ and\
  \bibinfo {author} {\bibfnamefont {A.}~\bibnamefont {{Vuorinen}}},\ }\href
  {https://doi.org/10.1038/s41567-020-0914-9} {\bibfield  {journal} {\bibinfo
  {journal} {Nature Physics}\ }\textbf {\bibinfo {volume} {16}},\ \bibinfo
  {pages} {907} (\bibinfo {year} {2020})}\BibitemShut {NoStop}%
\bibitem [{\citenamefont {Miller}\ \emph {et~al.}(2021)\citenamefont {Miller}
  \emph {et~al.}}]{Miller2021}%
  \BibitemOpen
  \bibfield  {author} {\bibinfo {author} {\bibfnamefont {M.}~\bibnamefont
  {Miller}} \emph {et~al.},\ }\href {https://doi.org/10.3847/2041-8213/ac089b}
  {\bibfield  {journal} {\bibinfo  {journal} {The Astrophysical Journal
  Letters}\ }\textbf {\bibinfo {volume} {918}},\ \bibinfo {pages} {L28}
  (\bibinfo {year} {2021})}\BibitemShut {NoStop}%
\bibitem [{\citenamefont {Riley}\ \emph {et~al.}(2021)\citenamefont {Riley}
  \emph {et~al.}}]{Riley2021}%
  \BibitemOpen
  \bibfield  {author} {\bibinfo {author} {\bibfnamefont {T.}~\bibnamefont
  {Riley}} \emph {et~al.},\ }\href {https://doi.org/10.3847/2041-8213/ac0a81}
  {\bibfield  {journal} {\bibinfo  {journal} {The Astrophysical Journal
  Letters}\ }\textbf {\bibinfo {volume} {918}},\ \bibinfo {pages} {L27}
  (\bibinfo {year} {2021})}\BibitemShut {NoStop}%
\bibitem [{\citenamefont {Romani}\ \emph {et~al.}(2022)\citenamefont {Romani},
  \citenamefont {Kandel}, \citenamefont {Filippenko}, \citenamefont {Brink},\
  and\ \citenamefont {Zheng}}]{romani}%
  \BibitemOpen
  \bibfield  {author} {\bibinfo {author} {\bibfnamefont {R.~W.}\ \bibnamefont
  {Romani}}, \bibinfo {author} {\bibfnamefont {D.}~\bibnamefont {Kandel}},
  \bibinfo {author} {\bibfnamefont {A.~V.}\ \bibnamefont {Filippenko}},
  \bibinfo {author} {\bibfnamefont {T.~G.}\ \bibnamefont {Brink}},\ and\
  \bibinfo {author} {\bibfnamefont {W.}~\bibnamefont {Zheng}},\ }\href
  {https://doi.org/10.3847/2041-8213/ac8007} {\bibfield  {journal} {\bibinfo
  {journal} {The Astrophysical Journal Letters}\ }\textbf {\bibinfo {volume}
  {934}},\ \bibinfo {pages} {L17} (\bibinfo {year} {2022})}\BibitemShut
  {NoStop}%
\bibitem [{\citenamefont {{Ivanenko}}\ and\ \citenamefont
  {{Kurdgelaidze}}(1965)}]{Ivanenko}%
  \BibitemOpen
  \bibfield  {author} {\bibinfo {author} {\bibfnamefont {D.~D.}\ \bibnamefont
  {{Ivanenko}}}\ and\ \bibinfo {author} {\bibfnamefont {D.~F.}\ \bibnamefont
  {{Kurdgelaidze}}},\ }\href {https://doi.org/10.1007/BF01042830} {\bibfield
  {journal} {\bibinfo  {journal} {Astrophysics}\ }\textbf {\bibinfo {volume}
  {1}},\ \bibinfo {pages} {251} (\bibinfo {year} {1965})}\BibitemShut {NoStop}%
\bibitem [{\citenamefont {Menezes}(2021)}]{debora-universe}%
  \BibitemOpen
  \bibfield  {author} {\bibinfo {author} {\bibfnamefont {D.~P.}\ \bibnamefont
  {Menezes}},\ }\href {https://www.mdpi.com/2218-1997/7/8/267} {\bibfield
  {journal} {\bibinfo  {journal} {Universe}\ }\textbf {\bibinfo {volume} {7}}
  (\bibinfo {year} {2021})}\BibitemShut {NoStop}%
\bibitem [{\citenamefont {Lalazissis}\ \emph {et~al.}(2009)\citenamefont
  {Lalazissis}, \citenamefont {Karatzikos}, \citenamefont {Fossion},
  \citenamefont {Arteaga}, \citenamefont {Afanasjev},\ and\ \citenamefont
  {Ring}}]{snl3}%
  \BibitemOpen
  \bibfield  {author} {\bibinfo {author} {\bibfnamefont {G.}~\bibnamefont
  {Lalazissis}}, \bibinfo {author} {\bibfnamefont {S.}~\bibnamefont
  {Karatzikos}}, \bibinfo {author} {\bibfnamefont {R.}~\bibnamefont {Fossion}},
  \bibinfo {author} {\bibfnamefont {D.~P.}\ \bibnamefont {Arteaga}}, \bibinfo
  {author} {\bibfnamefont {A.}~\bibnamefont {Afanasjev}},\ and\ \bibinfo
  {author} {\bibfnamefont {P.}~\bibnamefont {Ring}},\ }\href
  {https://doi.org/https://doi.org/10.1016/j.physletb.2008.11.070} {\bibfield
  {journal} {\bibinfo  {journal} {Physics Letters B}\ }\textbf {\bibinfo
  {volume} {671}},\ \bibinfo {pages} {36} (\bibinfo {year} {2009})}\BibitemShut
  {NoStop}%
\bibitem [{\citenamefont {Lopes}\ \emph {et~al.}(2021)\citenamefont {Lopes},
  \citenamefont {Biesdorf},\ and\ \citenamefont {Menezes}}]{Lopes_2021}%
  \BibitemOpen
  \bibfield  {author} {\bibinfo {author} {\bibfnamefont {L.~L.}\ \bibnamefont
  {Lopes}}, \bibinfo {author} {\bibfnamefont {C.}~\bibnamefont {Biesdorf}},\
  and\ \bibinfo {author} {\bibfnamefont {D.~P.}\ \bibnamefont {Menezes}},\
  }\href {https://doi.org/10.1088/1402-4896/abef34} {\bibfield  {journal}
  {\bibinfo  {journal} {Physica Scripta}\ }\textbf {\bibinfo {volume} {96}},\
  \bibinfo {pages} {065303} (\bibinfo {year} {2021})}\BibitemShut {NoStop}%
\bibitem [{\citenamefont {Panotopoulos}\ and\ \citenamefont
  {Lopes}(2017{\natexlab{b}})}]{ilidio}%
  \BibitemOpen
  \bibfield  {author} {\bibinfo {author} {\bibfnamefont {G.}~\bibnamefont
  {Panotopoulos}}\ and\ \bibinfo {author} {\bibfnamefont {I.}~\bibnamefont
  {Lopes}},\ }\href {https://doi.org/10.1103/PhysRevD.96.083004} {\bibfield
  {journal} {\bibinfo  {journal} {Phys. Rev. D}\ }\textbf {\bibinfo {volume}
  {96}},\ \bibinfo {pages} {083004} (\bibinfo {year}
  {2017}{\natexlab{b}})}\BibitemShut {NoStop}%
\bibitem [{\citenamefont {Cline}\ \emph {et~al.}(2013)\citenamefont {Cline},
  \citenamefont {Scott}, \citenamefont {Kainulainen},\ and\ \citenamefont
  {Weniger}}]{cline}%
  \BibitemOpen
  \bibfield  {author} {\bibinfo {author} {\bibfnamefont {J.~M.}\ \bibnamefont
  {Cline}}, \bibinfo {author} {\bibfnamefont {P.}~\bibnamefont {Scott}},
  \bibinfo {author} {\bibfnamefont {K.}~\bibnamefont {Kainulainen}},\ and\
  \bibinfo {author} {\bibfnamefont {C.}~\bibnamefont {Weniger}},\ }\href
  {https://doi.org/10.1103/PhysRevD.88.055025} {\bibfield  {journal} {\bibinfo
  {journal} {Phys. Rev. D}\ }\textbf {\bibinfo {volume} {88}},\ \bibinfo
  {pages} {055025} (\bibinfo {year} {2013})}\BibitemShut {NoStop}%
\bibitem [{\citenamefont {Cline}\ \emph {et~al.}(2015)\citenamefont {Cline},
  \citenamefont {Kainulainen}, \citenamefont {Scott},\ and\ \citenamefont
  {Weniger}}]{cline-errata}%
  \BibitemOpen
  \bibfield  {author} {\bibinfo {author} {\bibfnamefont {J.~M.}\ \bibnamefont
  {Cline}}, \bibinfo {author} {\bibfnamefont {K.}~\bibnamefont {Kainulainen}},
  \bibinfo {author} {\bibfnamefont {P.}~\bibnamefont {Scott}},\ and\ \bibinfo
  {author} {\bibfnamefont {C.}~\bibnamefont {Weniger}},\ }\href
  {https://doi.org/10.1103/PhysRevD.92.039906} {\bibfield  {journal} {\bibinfo
  {journal} {Phys. Rev. D}\ }\textbf {\bibinfo {volume} {92}},\ \bibinfo
  {pages} {039906} (\bibinfo {year} {2015})}\BibitemShut {NoStop}%
\bibitem [{\citenamefont {Feng}(2010)}]{cand1}%
  \BibitemOpen
  \bibfield  {author} {\bibinfo {author} {\bibfnamefont {J.~L.}\ \bibnamefont
  {Feng}},\ }\href {https://doi.org/10.1146/annurev-astro-082708-101659}
  {\bibfield  {journal} {\bibinfo  {journal} {Annual Review of Astronomy and
  Astrophysics}\ }\textbf {\bibinfo {volume} {48}},\ \bibinfo {pages} {495}
  (\bibinfo {year} {2010})}\BibitemShut {NoStop}%
\bibitem [{\citenamefont {Kusenko}\ and\ \citenamefont
  {Rosenberg}(2013)}]{cand2}%
  \BibitemOpen
  \bibfield  {author} {\bibinfo {author} {\bibfnamefont {A.}~\bibnamefont
  {Kusenko}}\ and\ \bibinfo {author} {\bibfnamefont {L.~J.}\ \bibnamefont
  {Rosenberg}},\ }\href {https://doi.org/10.48550/ARXIV.1310.8642} {\bibinfo
  {title} {Snowmass-2013 cosmic frontier 3 (cf3) working group summary:
  Non-wimp dark matter}} (\bibinfo {year} {2013})\BibitemShut {NoStop}%
\bibitem [{\citenamefont {Carlson}\ \emph {et~al.}(2022)\citenamefont
  {Carlson}, \citenamefont {Dutra}, \citenamefont {Lourenço},\ and\
  \citenamefont {Margueron}}]{brett-jerome}%
  \BibitemOpen
  \bibfield  {author} {\bibinfo {author} {\bibfnamefont {B.~V.}\ \bibnamefont
  {Carlson}}, \bibinfo {author} {\bibfnamefont {M.}~\bibnamefont {Dutra}},
  \bibinfo {author} {\bibfnamefont {O.}~\bibnamefont {Lourenço}},\ and\
  \bibinfo {author} {\bibfnamefont {J.}~\bibnamefont {Margueron}},\ }\href
  {https://doi.org/10.48550/ARXIV.2209.03257} {\bibinfo {title} {Low-energy
  nuclear physics and global neutron star properties}} (\bibinfo {year}
  {2022})\BibitemShut {NoStop}%
\bibitem [{\citenamefont {Lopes}\ \emph {et~al.}(2022)\citenamefont {Lopes},
  \citenamefont {Biesdorf},\ and\ \citenamefont {Menezes}}]{Lopes_2022}%
  \BibitemOpen
  \bibfield  {author} {\bibinfo {author} {\bibfnamefont {L.~L.}\ \bibnamefont
  {Lopes}}, \bibinfo {author} {\bibfnamefont {C.}~\bibnamefont {Biesdorf}},\
  and\ \bibinfo {author} {\bibfnamefont {D.~P.}\ \bibnamefont {Menezes}},\
  }\href {https://doi.org/10.1093/mnras/stac793} {\bibfield  {journal}
  {\bibinfo  {journal} {Monthly Notices of the Royal Astronomical Society}\
  }\textbf {\bibinfo {volume} {512}},\ \bibinfo {pages} {5110} (\bibinfo {year}
  {2022})}\BibitemShut {NoStop}%
\bibitem [{\citenamefont {Serot}(1992)}]{Serot_1992}%
  \BibitemOpen
  \bibfield  {author} {\bibinfo {author} {\bibfnamefont {B.~D.}\ \bibnamefont
  {Serot}},\ }\href {https://doi.org/10.1088/0034-4885/55/11/001} {\bibfield
  {journal} {\bibinfo  {journal} {Reports on Progress in Physics}\ }\textbf
  {\bibinfo {volume} {55}},\ \bibinfo {pages} {1855} (\bibinfo {year}
  {1992})}\BibitemShut {NoStop}%
\bibitem [{\citenamefont {Tolman}(1939)}]{tov39}%
  \BibitemOpen
  \bibfield  {author} {\bibinfo {author} {\bibfnamefont {R.~C.}\ \bibnamefont
  {Tolman}},\ }\href {https://doi.org/10.1103/PhysRev.55.364} {\bibfield
  {journal} {\bibinfo  {journal} {Phys. Rev.}\ }\textbf {\bibinfo {volume}
  {55}},\ \bibinfo {pages} {364} (\bibinfo {year} {1939})}\BibitemShut
  {NoStop}%
\bibitem [{\citenamefont {Oppenheimer}\ and\ \citenamefont
  {Volkoff}(1939)}]{tov39a}%
  \BibitemOpen
  \bibfield  {author} {\bibinfo {author} {\bibfnamefont {J.~R.}\ \bibnamefont
  {Oppenheimer}}\ and\ \bibinfo {author} {\bibfnamefont {G.~M.}\ \bibnamefont
  {Volkoff}},\ }\href {https://doi.org/10.1103/PhysRev.55.374} {\bibfield
  {journal} {\bibinfo  {journal} {Phys. Rev.}\ }\textbf {\bibinfo {volume}
  {55}},\ \bibinfo {pages} {374} (\bibinfo {year} {1939})}\BibitemShut
  {NoStop}%
\bibitem [{\citenamefont {{Gondek}}\ \emph {et~al.}(1997)\citenamefont
  {{Gondek}}, \citenamefont {{Haensel}},\ and\ \citenamefont
  {{Zdunik}}}]{rad1}%
  \BibitemOpen
  \bibfield  {author} {\bibinfo {author} {\bibfnamefont {D.}~\bibnamefont
  {{Gondek}}}, \bibinfo {author} {\bibfnamefont {P.}~\bibnamefont
  {{Haensel}}},\ and\ \bibinfo {author} {\bibfnamefont {J.~L.}\ \bibnamefont
  {{Zdunik}}},\ }\href {https://ui.adsabs.harvard.edu/abs/1997A&A...325..217G}
  {\bibfield  {journal} {\bibinfo  {journal} {Astronomy and Astrophysics}\
  }\textbf {\bibinfo {volume} {325}},\ \bibinfo {pages} {217} (\bibinfo {year}
  {1997})}\BibitemShut {NoStop}%
\bibitem [{\citenamefont {Parisi}\ \emph {et~al.}(2021)\citenamefont {Parisi},
  \citenamefont {Flores}, \citenamefont {Lenzi}, \citenamefont {Chen},\ and\
  \citenamefont {Lugones}}]{rad2}%
  \BibitemOpen
  \bibfield  {author} {\bibinfo {author} {\bibfnamefont {A.}~\bibnamefont
  {Parisi}}, \bibinfo {author} {\bibfnamefont {C.~V.}\ \bibnamefont {Flores}},
  \bibinfo {author} {\bibfnamefont {C.~H.}\ \bibnamefont {Lenzi}}, \bibinfo
  {author} {\bibfnamefont {C.-S.}\ \bibnamefont {Chen}},\ and\ \bibinfo
  {author} {\bibfnamefont {G.}~\bibnamefont {Lugones}},\ }\href
  {https://doi.org/10.1088/1475-7516/2021/06/042} {\bibfield  {journal}
  {\bibinfo  {journal} {Journal of Cosmology and Astroparticle Physics}\
  }\textbf {\bibinfo {volume} {2021}}\bibinfo  {number} { (06)},\ \bibinfo
  {pages} {042}}\BibitemShut {NoStop}%
\bibitem [{\citenamefont {Pereira}\ \emph {et~al.}(2018)\citenamefont
  {Pereira}, \citenamefont {Flores},\ and\ \citenamefont {Lugones}}]{rad3}%
  \BibitemOpen
\bibfield  {number} {  }\bibfield  {author} {\bibinfo {author} {\bibfnamefont
  {J.~P.}\ \bibnamefont {Pereira}}, \bibinfo {author} {\bibfnamefont {C.~V.}\
  \bibnamefont {Flores}},\ and\ \bibinfo {author} {\bibfnamefont
  {G.}~\bibnamefont {Lugones}},\ }\href
  {https://doi.org/10.3847/1538-4357/aabfbf} {\bibfield  {journal} {\bibinfo
  {journal} {The Astrophysical Journal}\ }\textbf {\bibinfo {volume} {860}},\
  \bibinfo {pages} {12} (\bibinfo {year} {2018})}\BibitemShut {NoStop}%
\bibitem [{\citenamefont {Mariani}\ \emph {et~al.}(2019)\citenamefont
  {Mariani}, \citenamefont {Orsaria}, \citenamefont {Ranea-Sandoval},\ and\
  \citenamefont {Lugones}}]{rad4}%
  \BibitemOpen
  \bibfield  {author} {\bibinfo {author} {\bibfnamefont {M.}~\bibnamefont
  {Mariani}}, \bibinfo {author} {\bibfnamefont {M.~G.}\ \bibnamefont
  {Orsaria}}, \bibinfo {author} {\bibfnamefont {I.~F.}\ \bibnamefont
  {Ranea-Sandoval}},\ and\ \bibinfo {author} {\bibfnamefont {G.}~\bibnamefont
  {Lugones}},\ }\href {https://doi.org/10.1093/mnras/stz2392} {\bibfield
  {journal} {\bibinfo  {journal} {Monthly Notices of the Royal Astronomical
  Society}\ }\textbf {\bibinfo {volume} {489}},\ \bibinfo {pages} {4261}
  (\bibinfo {year} {2019})}\BibitemShut {NoStop}%
\bibitem [{\citenamefont {Arba\~nil}\ and\ \citenamefont
  {Malheiro}(2015)}]{rad5}%
  \BibitemOpen
  \bibfield  {author} {\bibinfo {author} {\bibfnamefont {J.~D.~V.}\
  \bibnamefont {Arba\~nil}}\ and\ \bibinfo {author} {\bibfnamefont
  {M.}~\bibnamefont {Malheiro}},\ }\href
  {https://doi.org/10.1103/PhysRevD.92.084009} {\bibfield  {journal} {\bibinfo
  {journal} {Phys. Rev. D}\ }\textbf {\bibinfo {volume} {92}},\ \bibinfo
  {pages} {084009} (\bibinfo {year} {2015})}\BibitemShut {NoStop}%
\bibitem [{\citenamefont {Sun}\ \emph {et~al.}(2021)\citenamefont {Sun},
  \citenamefont {Zheng}, \citenamefont {Chen}, \citenamefont {Burgio},\ and\
  \citenamefont {Schulze}}]{rad6}%
  \BibitemOpen
  \bibfield  {author} {\bibinfo {author} {\bibfnamefont {T.-T.}\ \bibnamefont
  {Sun}}, \bibinfo {author} {\bibfnamefont {Z.-Y.}\ \bibnamefont {Zheng}},
  \bibinfo {author} {\bibfnamefont {H.}~\bibnamefont {Chen}}, \bibinfo {author}
  {\bibfnamefont {G.~F.}\ \bibnamefont {Burgio}},\ and\ \bibinfo {author}
  {\bibfnamefont {H.-J.}\ \bibnamefont {Schulze}},\ }\href
  {https://doi.org/10.1103/PhysRevD.103.103003} {\bibfield  {journal} {\bibinfo
   {journal} {Phys. Rev. D}\ }\textbf {\bibinfo {volume} {103}},\ \bibinfo
  {pages} {103003} (\bibinfo {year} {2021})}\BibitemShut {NoStop}%
\bibitem [{\citenamefont {Jim\'enez}\ and\ \citenamefont {Fraga}(2019)}]{rad7}%
  \BibitemOpen
  \bibfield  {author} {\bibinfo {author} {\bibfnamefont {J.~C.}\ \bibnamefont
  {Jim\'enez}}\ and\ \bibinfo {author} {\bibfnamefont {E.~S.}\ \bibnamefont
  {Fraga}},\ }\href {https://doi.org/10.1103/PhysRevD.100.114041} {\bibfield
  {journal} {\bibinfo  {journal} {Phys. Rev. D}\ }\textbf {\bibinfo {volume}
  {100}},\ \bibinfo {pages} {114041} (\bibinfo {year} {2019})}\BibitemShut
  {NoStop}%
\bibitem [{\citenamefont {{Haensel}}\ \emph {et~al.}(1989)\citenamefont
  {{Haensel}}, \citenamefont {{Zdunik}},\ and\ \citenamefont
  {{Schaeffer}}}]{rad0}%
  \BibitemOpen
  \bibfield  {author} {\bibinfo {author} {\bibfnamefont {P.}~\bibnamefont
  {{Haensel}}}, \bibinfo {author} {\bibfnamefont {J.~L.}\ \bibnamefont
  {{Zdunik}}},\ and\ \bibinfo {author} {\bibfnamefont {R.}~\bibnamefont
  {{Schaeffer}}},\ }\href
  {https://ui.adsabs.harvard.edu/abs/1989A&A...217..137H} {\bibfield  {journal}
  {\bibinfo  {journal} {Astronomy and Astrophysics}\ }\textbf {\bibinfo
  {volume} {217}},\ \bibinfo {pages} {137} (\bibinfo {year}
  {1989})}\BibitemShut {NoStop}%
\bibitem [{\citenamefont {Riley}\ \emph {et~al.}(2019)\citenamefont {Riley}
  \emph {et~al.}}]{Riley:2019yda}%
  \BibitemOpen
  \bibfield  {author} {\bibinfo {author} {\bibfnamefont {T.~E.}\ \bibnamefont
  {Riley}} \emph {et~al.},\ }\href {https://doi.org/10.3847/2041-8213/ab481c}
  {\bibfield  {journal} {\bibinfo  {journal} {Astrophys. J. Lett.}\ }\textbf
  {\bibinfo {volume} {887}},\ \bibinfo {pages} {L21} (\bibinfo {year}
  {2019})}\BibitemShut {NoStop}%
\bibitem [{\citenamefont {Miller}\ \emph {et~al.}(2019)\citenamefont {Miller}
  \emph {et~al.}}]{Miller:2019cac}%
  \BibitemOpen
  \bibfield  {author} {\bibinfo {author} {\bibfnamefont {M.}~\bibnamefont
  {Miller}} \emph {et~al.},\ }\href {https://doi.org/10.3847/2041-8213/ab50c5}
  {\bibfield  {journal} {\bibinfo  {journal} {Astrophys. J. Lett.}\ }\textbf
  {\bibinfo {volume} {887}},\ \bibinfo {pages} {L24} (\bibinfo {year}
  {2019})}\BibitemShut {NoStop}%
\bibitem [{\citenamefont {Abbott}\ \emph {et~al.}(2019)\citenamefont {Abbott}
  \emph {et~al.}}]{Abbott:2018wiz}%
  \BibitemOpen
  \bibfield  {author} {\bibinfo {author} {\bibfnamefont {B.}~\bibnamefont
  {Abbott}} \emph {et~al.},\ }\href {https://doi.org/10.1103/PhysRevX.9.011001}
  {\bibfield  {journal} {\bibinfo  {journal} {Phys. Rev. X}\ }\textbf {\bibinfo
  {volume} {9}},\ \bibinfo {pages} {011001} (\bibinfo {year}
  {2019})}\BibitemShut {NoStop}%
\bibitem [{\citenamefont {Louren\c{c}o}\ \emph {et~al.}(2021)\citenamefont
  {Louren\c{c}o}, \citenamefont {Lenzi}, \citenamefont {Dutra}, \citenamefont
  {Ferrer}, \citenamefont {de~la Incera}, \citenamefont {Paulucci},\ and\
  \citenamefont {Horvath}}]{ferrer}%
  \BibitemOpen
  \bibfield  {author} {\bibinfo {author} {\bibfnamefont {O.}~\bibnamefont
  {Louren\c{c}o}}, \bibinfo {author} {\bibfnamefont {C.~H.}\ \bibnamefont
  {Lenzi}}, \bibinfo {author} {\bibfnamefont {M.}~\bibnamefont {Dutra}},
  \bibinfo {author} {\bibfnamefont {E.~J.}\ \bibnamefont {Ferrer}}, \bibinfo
  {author} {\bibfnamefont {V.}~\bibnamefont {de~la Incera}}, \bibinfo {author}
  {\bibfnamefont {L.}~\bibnamefont {Paulucci}},\ and\ \bibinfo {author}
  {\bibfnamefont {J.~E.}\ \bibnamefont {Horvath}},\ }\href
  {https://doi.org/10.1103/PhysRevD.103.103010} {\bibfield  {journal} {\bibinfo
   {journal} {Phys. Rev. D}\ }\textbf {\bibinfo {volume} {103}},\ \bibinfo
  {pages} {103010} (\bibinfo {year} {2021})}\BibitemShut {NoStop}%
\bibitem [{\citenamefont {Chatziioannou}\ \emph {et~al.}(2018)\citenamefont
  {Chatziioannou}, \citenamefont {Haster},\ and\ \citenamefont
  {Zimmerman}}]{chatziioannou}%
  \BibitemOpen
  \bibfield  {author} {\bibinfo {author} {\bibfnamefont {K.}~\bibnamefont
  {Chatziioannou}}, \bibinfo {author} {\bibfnamefont {C.-J.}\ \bibnamefont
  {Haster}},\ and\ \bibinfo {author} {\bibfnamefont {A.}~\bibnamefont
  {Zimmerman}},\ }\href {https://doi.org/10.1103/PhysRevD.97.104036} {\bibfield
   {journal} {\bibinfo  {journal} {Phys. Rev. D}\ }\textbf {\bibinfo {volume}
  {97}},\ \bibinfo {pages} {104036} (\bibinfo {year} {2018})}\BibitemShut
  {NoStop}%
\bibitem [{\citenamefont {Flores}\ \emph {et~al.}(2020)\citenamefont {Flores},
  \citenamefont {Lopes}, \citenamefont {Benito},\ and\ \citenamefont
  {Menezes}}]{Flores2020}%
  \BibitemOpen
  \bibfield  {author} {\bibinfo {author} {\bibfnamefont {C.}~\bibnamefont
  {Flores}}, \bibinfo {author} {\bibfnamefont {L.}~\bibnamefont {Lopes}},
  \bibinfo {author} {\bibfnamefont {L.}~\bibnamefont {Benito}},\ and\ \bibinfo
  {author} {\bibfnamefont {D.}~\bibnamefont {Menezes}},\ }\href
  {https://doi.org/10.1140/epjc/s10052-020-08705-1} {\bibfield  {journal}
  {\bibinfo  {journal} {The European Physical Journal C}\ }\textbf {\bibinfo
  {volume} {80}},\ \bibinfo {pages} {1142} (\bibinfo {year}
  {2020})}\BibitemShut {NoStop}%
\bibitem [{\citenamefont {Takátsy}\ and\ \citenamefont
  {Kovács}(2020)}]{takatsy}%
  \BibitemOpen
  \bibfield  {author} {\bibinfo {author} {\bibfnamefont {J.}~\bibnamefont
  {Takátsy}}\ and\ \bibinfo {author} {\bibfnamefont {P.}~\bibnamefont
  {Kovács}},\ }\href@noop {} {\bibfield  {journal} {\bibinfo  {journal} {Phys.
  Rev. D}\ }\textbf {\bibinfo {volume} {102}},\ \bibinfo {pages} {028501}
  (\bibinfo {year} {2020})}\BibitemShut {NoStop}%
\bibitem [{\citenamefont {Sen}\ and\ \citenamefont {Guha}(2021)}]{feebly}%
  \BibitemOpen
  \bibfield  {author} {\bibinfo {author} {\bibfnamefont {D.}~\bibnamefont
  {Sen}}\ and\ \bibinfo {author} {\bibfnamefont {A.}~\bibnamefont {Guha}},\
  }\href {https://doi.org/10.1093/mnras/stab1056} {\bibfield  {journal}
  {\bibinfo  {journal} {Monthly Notices of the Royal Astronomical Society}\
  }\textbf {\bibinfo {volume} {504}},\ \bibinfo {pages} {3354} (\bibinfo {year}
  {2021})}\BibitemShut {NoStop}%
\end{thebibliography}%
\bibliographystyle{apsrev4-2}

\end{document}